\documentclass[iop,numberedappendix]{emulateapj-rtx4}

\usepackage{mathrsfs }
\usepackage{natbib}
\usepackage{multirow}
\usepackage{amssymb}
\usepackage{apjfonts}
\usepackage{ulem} 
\newcommand{\NHI}{\ensuremath{N_{\rm HI}}}
\newcommand{\fnx}{\ensuremath{f(N_{\rm HI}, X)}}
\newcommand{\mfp}{\ensuremath{\lambda_{\rm mfp}}}
\newcommand{\mfps}{\ensuremath{\lambda_{\rm{mfp}}^{\rm{static}}}}
\newcommand{\mfpsIGM}{\ensuremath{\lambda_{\rm{mfp,~IGM}}^{\rm{static}}}}
\newcommand{\mfpsCGM}{\ensuremath{\lambda_{\rm{mfp,~IGM+CGM}}^{\rm{static}}}}

\def\kms{km~s$^{-1}$}
\def\cm2{cm$^{-2}$}
\def\Msun{$M_{\odot}$}

\newcommand{\bd}{\ensuremath{{b_{\rm d}}}}

\def\ltsima{$\; \buildrel < \over \sim \;$}
\def\gtsima{$\; \buildrel > \over \sim \;$}
\def\simgt{\lower.5ex\hbox{\gtsima}}
\def\simlt{\lower.5ex\hbox{\ltsima}}

\usepackage{float}

\begin{document}

\title{The Column Density Distribution and Continuum Opacity of \\ the Intergalactic 
and Circumgalactic Medium at redshift $\langle z \rangle =2.4$}
\author{
 Gwen C. Rudie,\altaffilmark{2}
 Charles C. Steidel,\altaffilmark{2}
 Alice E. Shapley,\altaffilmark{3} \&
Max Pettini\altaffilmark{4} 
 }

\altaffiltext{1}{Based on data obtained at the W.M. Keck Observatory, which is operated as a scientific partnership among the California Institute of Technology, the University of California,  and the National Aeronautics and Space Administration, and was made possible by the generous financial support of the W.M. Keck Foundation.}
\altaffiltext{2}{Cahill Center for Astronomy and Astrophysics, California Institute of Technology, MS 249-17, Pasadena, CA 91125, USA}
\altaffiltext{3}{Department of Astronomy, University of California, Los Angeles, 430 Portola Plaza, Los Angeles, CA 90024, USA}
\altaffiltext{4}{Institute of Astronomy, Madingley Road, Cambridge CB3 0HA, UK}


\email{gwen@astro.caltech.edu}


\shortauthors{Rudie et~al.}


\shorttitle{IGM and CGM Opacity at $\langle z \rangle =2.4$}

\begin{abstract}
We present new high-precision measurements of the opacity of the intergalactic and circumgalactic medium (IGM, CGM) at  $\langle z \rangle =2.4$. Using Voigt profile fits to the full Lyman $\alpha$ and Lyman $\beta$ forests in 15 high-resolution high-S/N spectra of hyperluminous QSOs, we make the first statistically robust measurement of the frequency of absorbers with \ion{H}{1} column densities $14 \lesssim \log(\NHI/ \rm{cm}^{-2}) \lesssim 17.2$.  We also present the first measurements of the frequency distribution of \ion{H}{1} absorbers in the volume surrounding high-$z$ galaxies (the CGM, 300 pkpc), finding that the incidence of absorbers in the CGM is much higher than in the IGM. In agreement with \citet{gcr12}, we find that there are fractionally more high-\NHI\ absorbers than low-\NHI\ absorbers in the CGM compared to the IGM, leading to a shallower power law fit to the CGM frequency distribution.  We use these new measurements to calculate the total opacity of the IGM and CGM to hydrogen-ionizing photons, finding significantly higher opacity than most previous studies, especially from absorbers with $\log(\NHI/ \rm{cm}^{-2}) < 17.2$. Reproducing the opacity measured in our data as well as the incidence of absorbers with $\log(\NHI/ \rm{cm}^{-2}) > 17.2$ requires a broken power law parameterization of the frequency distribution with a break near \NHI $\approx10^{15}$ \cm2. We compute new estimates of the mean free path (\mfp) to hydrogen-ionizing photons at $z_{\rm em} = 2.4$, finding $\mfp = 147 \pm 15$ Mpc when considering only IGM opacity. If instead, we consider photons emanating from a high-$z$ star-forming galaxy and account for the local excess opacity due to the surrounding CGM of the galaxy itself, the mean free path is reduced to $\mfp = 121 \pm 15$ Mpc. These \mfp\ measurements are smaller than recent estimates and should inform future studies of the metagalactic UV background and of ionizing sources at $z\approx2-3$.
\end{abstract}


\section{Introduction}

\label{intro}

Observations of the Lyman $\alpha$ forest imprinted upon the spectra of background QSOs provide a rare window into the nature and evolution of baryonic structures in the universe. With the advent of the theory of cold dark matter (CDM), it was recognized that self-gravitating halos could be the sites of uncollapsed Ly$\alpha$ clouds \citep{ume85,ree86}. 
In the modern version of this model, the Ly$\alpha$ forest simply represents the array of density fluctuations present in the early universe and evolved under gravity to the present day \citep{tyt87,cen94,her96}, modulo the effect of the formation of stars, black holes, and galaxies which appear to have polluted a large fraction of the cosmic hydrogen with metals, even at early times \citep{son96, dav98,sch03,sim04,rya09,bec09,sim11,sim11b,bec11b,bec12} and led to the ionization of the vast majority of the gas \citep{gun65}.  

There is an extensive literature concerned with the measurement of cosmological parameters using observations of the Ly$\alpha$ forest [for a review, see \citet{rau98} and \citet{mei09}]. Considerable progress has been made in the last several decades in measuring the detailed properties of the IGM and their implications for cosmology, structure formation, reionization, and the metagalactic background. 

One of the most fundamental of these measurements is the number of absorbers as a function of column density (\NHI) and redshift that can be seen along a given path through the Universe. 
This measurement has required significant observational effort, in a large part due to the very large dynamic range of the measurements ($10^{12} \lesssim \NHI \lesssim 10^{22}$ \cm2) which have necessitated surveys optimized for particular ranges in \NHI. 
This requirement is easily understood by considering the curve of growth of the Ly$\alpha$ line of neutral hydrogen.

 The most common absorbers with $\log(\NHI/ \rm{cm}^{-2}) \lesssim14$ are optically thin in the Lyman $\alpha$ transition and are thus on the linear part of the curve of growth. As such, their properties can be measured with a few high-resolution ($\sim$10 \kms) high-signal to noise ratio (S/N $\sim$ 50) spectra \citep{kim02}. At the highest column densities, $\log(\NHI/ \rm{cm}^{-2}) \gtrsim 20.3$, the Lorentzian wings of the absorption line in Ly$\alpha$ become obvious against the continuum of the background source. Such rare pockets of predominantly neutral hydrogen, commonly referred to as Damped Lyman $\alpha$ absorbers (DLAs), can be identified and measured to high statistical precision with a large number of low to moderate resolution spectra of background QSOs as has been done using (e.g.) the QSO sample within the Sloan Digital Sky Survey \citep{not09}. 
 
 Absorbers with intermediate \NHI, $14 \lesssim \log(\NHI/ \rm cm^{-2}) \lesssim 20$, lie on the flat part of the curve of growth thus requiring different techniques. Systems with \NHI $> 10^{17.2}$ cm$^{-2}$ have an optical depth to hydrogen-ionizing photons $\tau \ge 1$, and so may be recognized by the strong breaks they produce in background QSO spectra at the Lyman limit (912\AA) in their rest frame. These so-called Lyman Limit Systems (LLS) can thus be easily discovered using low resolution spectra; however, a large sample is needed because they are comparatively rare. For studies at $z\lesssim2.6$, surveys of LLSs require space-based observations as the Lyman limit shifts below the atmospheric cutoff at UV wavelengths. However, over the last 30 years with a combination of ground and space based observations, the distribution of LLSs at $2<z<3$ has been well characterized \citep{tyt82,sar89,ste95,rib11,ome12}.
 
 Finally there is the historically most-problematic range, $14 \lesssim \log(\NHI/ \rm{cm}^{-2}) \lesssim 17.2$. These intermediate-\NHI\ absorbers are saturated in Ly$\alpha$ but have relatively low opacity to hydrogen-ionizing photons and therefore weak ``Lyman breaks.'' For this reason, observational constraints on their statistical incidence have been approximate at best.
 
In this paper, we address these intermediate-\NHI\ absorbers using Voigt profile fits to all-available higher-order Lyman series transitions. We take advantage of the increased dynamic range of the measurements afforded by the decreased oscillator strength of the Ly$\beta,\gamma,$ etc. transitions to accurately measure the frequency distribution of these absorbers, $f(N,X)$, for the first time in a statistically robust sample. In our derivation of an analytic representation of $f(N,X)$ we place emphasis on finding a method which best reproduces the measured opacity of absorbers to hydrogen ionizing sources.

These new measurements are crucial to our understanding of the opacity of the universe to hydrogen-ionizing photons. While individually absorbers with $\log(\NHI/ \rm{cm}^{-2}) < 17.2$ have relatively low optical depths at their Lyman limit, they are much more common than LLSs, and therefore their ensemble contributes nearly half of the IGM's opacity. 
Here, we present the most precise measurements of the continuum opacity of absorbers that are optically thin at the Lyman limit.


In this paper, we also attempt to account for the fact that galaxies are located in regions of relatively high gas density.  \citet{gcr12} clearly demonstrated that typical galaxies at $z\approx 2.3$ have large quantities of \ion{H}{1} surrounding them to 300 physical kpc (pkpc), and that there is excess \ion{H}{1} compared to random places in the IGM to $\sim$2 physical Mpc (pMpc).  In the case that galaxies such as those studied in \citet{gcr12} are significant contributors to the ionizing background, the effect of their surrounding circumgalactic gas distribution is an important consideration. In this work, we attempt to separate the probability that an ionizing photon will escape the ISM of its galaxy from the probability that such a photon, after escaping to 50 pkpc, will also escape the enhanced opacity found in its circumgalactic medium (CGM). 


We begin in Section \ref{data} with a description of the unique data that enable these very precise \NHI\ measurements. In Section \ref{fN_text} we present the new \NHI\ frequency distribution measurements. Section \ref{gal_fN} considers the frequency distribution within the CGM of galaxies and Section \ref{LyC} quantifies the total opacity of the IGM and the environment of the source galaxy. Updated measurements of the mean free path (\mfp) to hydrogen-ionizing photons are presented in Section \ref{mcText} along with their implications. The paper is summarized in Section \ref{con}.

Throughout this paper we assume a $\Lambda$-CDM cosmology with $H_{0} = 70$ \kms\  Mpc$^{-1}$, $\Omega_{\rm m} = 0.3$, and $\Omega_{\Lambda} = 0.7$. All distances are expressed in physical (proper) units unless stated otherwise. We use the abbreviations pkpc and pMpc to indicate physical units.
At the path-weighted mean redshift of the absorber sample ($\langle z \rangle = 2.37$), the age of the universe is 2.7 Gyr, the look-back time is 10.7 Gyr, and 8.2 pkpc subtends one arcsecond on the sky.

\section{Data and Analysis}

\label{data}

The data presented in this paper constitute a subset of the Keck Baryonic Structure Survey (KBSS). The KBSS was designed to detect and characterize the gaseous distribution surrounding star-forming galaxies at $2 < z < 3$ during the peak of cosmic star formation and black hole growth. The data include rest-frame UV spectra of 2188 star-forming galaxies located in 15 fields surrounding the lines of sight to hyper-luminous ($m_V \simeq 15.5-17$) QSOs. The present work focuses primarily on the \ion{H}{1} absorption line statistics resulting from analysis of high-resolution, high signal-to-noise echelle spectra of the 15 KBSS QSOs. A full description of the KBSS will be presented by Steidel et al. (in prep), but the portions of the data most relevant to this work (as well as a detailed description of the analysis of the QSO data) can be found in \citet{gcr12}. Here, we provide a brief description.

The 15 KBSS QSOs were observed primarily with the High Resolution Echelle Spectrometer \citep[HIRES;][]{vog94} on the Keck I telescope. The HIRES spectra have $R\simeq 45,000$ (FWHM$\simeq 7$ \kms), S/N $\sim 50-200$ per pixel, and cover at least the wavelength range 3100 -- 6000 \AA\ with no spectral gaps. The UV/blue wavelength coverage provides a significant advantage over other data sets as it enables the observation of Ly$\beta$ $\lambda 1025.7$ down to at least $z= 2.2$ in all 15 of our sightlines, and to significantly lower redshift in many.\footnote{The minimum wavelength that can be used for line fitting is determined by both the wavelength coverage of the spectrum (set by the spectrograph and the atmosphere) and by the location of the highest-redshift Lyman limit system along each line of sight. The minimum redshift for which we perform line fitting is shown in Table \ref{field}. }  The additional constraints provided by Ly$\beta$ (and in many cases, additional Lyman series transitions) allow for much more accurate measurements of \ion{H}{1} for \NHI$=10^{14}-10^{17} $ \cm2\ than can be made with spectra including only the Ly$\alpha$ transition. A full discussion of the redshift sensitivity of our fitting procedure is given in Appendix \ref{App_fitting}.

\begin{deluxetable*}{llllccccccc}
\tablecaption{KBSS Absorption Line Sample}  
\tablewidth{0pt}
\tablehead{
\colhead{Name} & \colhead{RA} & \colhead{Dec} & \colhead{$z_{\rm QSO}$\tablenotemark{a}} & \colhead{$\lambda_{\rm min}$\tablenotemark{b} [\AA]}  &  \colhead{$z$ range} & \colhead{$\Delta z$} & \colhead{$\Delta X$\tablenotemark{c}} & \colhead{N$_{\rm gal}$\tablenotemark{d}} & \colhead{S/N Ly$\alpha$\tablenotemark{e}} & \colhead{S/N Ly$\beta$\tablenotemark{e}}}
\startdata
 Q0100+130 (PHL957) & 01:03:11.3 &  $+$13:16:18.2  &  2.721 & 3133 &  2.0617-- 2.6838 &  0.62 &   2.02 & 5  & ~77   &   50  \\
 HS0105+1619 & 01:08:06.4  &  $+$16:35:50.0  &  2.652 & 3230 &  2.1561-- 2.6153 &  0.46 &   1.50  & 0 &127  	&   89  \\
 Q0142$-$09 (UM673a) & 01:45:16.6  &  $-$09:45:17.0  &  2.743 & 3097 &   2.0260-- 2.7060   &  0.68 & 2.21 &  3 & ~71  	&   45  \\
 Q0207$-$003 (UM402) & 02:09:50.7 &  $-$00:05:06.5  &  2.872 & 3227 & 2.1532-- 2.8339  &  0.68 &   2.26 &  1  &   ~82  	&   55  \\
 Q0449$-$1645 & 04:52:14.3  &  $-$16:40:16.2  &  2.684 &   3151 &  2.0792-- 2.6470  &  0.57 &    1.84 & 2  &  ~73  	&   41  \\
 Q0821+3107 & 08:21:07.6 &  $+$31:07:51.2 &  2.616 & 3239 &    2.1650-- 2.5794  &  0.41 &   1.35&  2  &   ~50  	&   33  \\
 Q1009+29 (CSO 38) & 10:11:55.6 &  $+$29:41:41.7  &  2.652\tablenotemark{f} &  3186 &   2.1132-- 2.6031	 &  0.49 &  1.59  & 0  &   ~99  	&   58  \\
 SBS1217+499 & 12:19:30.9 &  $+$49:40:51.2  &  2.704  &   3098 & 	 2.0273-- 2.6669     &  0.64 &  2.07  & 5 &  ~68  	&   38  \\
 HS1442+2931 & 14:44:53.7 &  $+$29:19:05.6  &  2.660 & 3152 &    2.0798-- 2.6237  &  0.54 &   1.76 & 6  &  ~99  	&   47  \\
 HS1549+1919 & 15:51:52.5  &  $+$19:11:04.3  &  2.843 &    3165 & 	 2.0926-- 2.8048  &  0.71 &   2.35 &  4 &  173  	&   74  \\
HS1603+3820 & 16:04:55.4 &  $+$38:12:01.8  &  2.551\tablenotemark{g} & 3181 &  2.1087-- 2.5066  &  0.40 &   1.28 &  2  &  108  	&   58  \\
 Q1623+268 (KP77) & 16:25:48.8 &  $+$26:46:58.8  &  2.535 &   3126 &  2.0544-- 2.4999  &  0.45 &  174 &   3   &    ~48  	&   28  \\
 HS1700+64 & 17:01:00.6  &  $+$64:12:09.4  &  2.751 &   3138 & 2.0668-- 2.7138  &  0.65 &    2.11 & 4   &     ~98  	&   42  \\
 Q2206$-$199 & 22:08:52.1  &  $-$19:43:59.7  &  2.573 &  3084 &  2.0133-- 2.5373   &  0.52 &    1.68 & 3  & 	   ~88  	&   46  \\
 Q2343+125 & 23:46:28.3 &  $+$12:48:57.8  &  2.573 &   3160 & 2.0884-- 2.5373 &  0.45 &    1.45 & 3   &                     ~71  	&   45      
 \enddata
 \tablenotetext{a}{The redshift of the QSO \citep{tra12}}
 \tablenotetext{b}{The minimum wavelength covered in the HIRES QSO spectrum}
\tablenotetext{c}{The comoving pathlength in this line of sight (equation \ref{pathlength}).}
 \tablenotetext{d}{The number of galaxies in our LRIS survey with impact parameters $<$ 300 pkpc.}
 \tablenotetext{e}{The average signal to noise ratio per pixel of the QSO spectrum in the wavelength range pertaining to IGM Ly$\alpha$ and Ly$\beta$ absorption.}
      \tablenotetext{f}{The redshift of this QSO was revised after the fitting of the HIRES spectrum was completed. The maximum redshift assumed $z_{\rm QSO}=2.640$}
       \tablenotetext{g}{The redshift of this QSO was revised after the fitting of the HIRES spectrum was completed. The maximum redshift assumed $z_{\rm QSO}=2.542$}
     \label{field}
\end{deluxetable*}

\begin{figure}
\center
\includegraphics[width=0.4\textwidth]{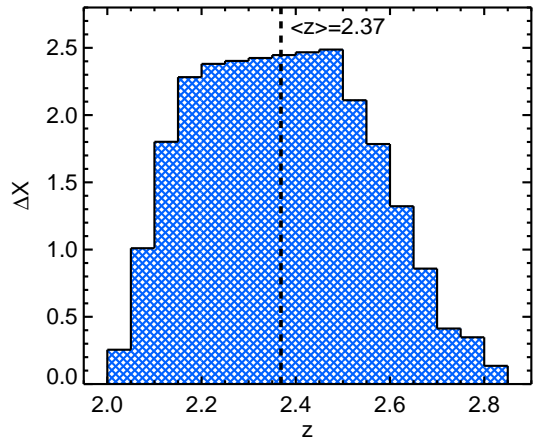}
\caption{\small Histogram of the pathlength sampled as a function of redshift. The path-length-weighted average redshift of the sample is $z=2.37$.}
\label{z_X}
\end{figure}

The reduction of the HIRES spectra was performed using T. Barlow's MAKEE package as described in detail by \citet{gcr12}. The analysis presented here includes a Voigt profile fit to the full Ly$\alpha$ and Ly$\beta$ forests in these spectra. The redshift range included in the fit is bounded at high redshift by the proximity zone of the QSO (taken to be 3000 \kms) in which absorbers may be ejected from or ionized by the QSO itself. The low-redshift cut was chosen to ensure that all \ion{H}{1} absorbers included in the sample had been observed in both the Ly$\alpha$ and Ly$\beta$ transition. The final redshift windows used for each spectrum are detailed in Table \ref{field}.

The line fitting was facilitated by a semi-automatic code that estimates input parameters (redshift $z$, \ion{H}{1} column density \NHI, and Doppler parameter \bd\ of each absorber) using a cross-correlation technique as described in \citet{gcr12}. These parameters were then input into the $\chi^2$ minimization code VPFIT\footnote{http://www.ast.cam.ac.uk/$\sim$rfc/vpfit.html; \copyright ~2007 R.F. Carswell, J.K. Webb} written by  R.F. Carswell and J.K. Webb. The results were iteratively checked, altered, and re-run until a good fit was achieved.  Median uncertainties in $\log(\NHI)$ for these absorber measurements are listed in Table \ref{NHI_error}. Examples of fits to absorbers with $\log(\NHI/\rm cm^{-2}) > 15.5$ are given in Appendix \ref{App_fitting} along with an assessment of the completeness of the catalog as a function of redshift. 

The result of this process is an \ion{H}{1} absorber catalog including 5758 absorbers with $12.0 < \log(\NHI/ \rm{cm}^{-2}) < 17.2$ and $ 2.02 < z <  2.84$ over a total redshift pathlength of $\Delta z = 8.27$. 
This represents the largest Ly$\alpha$ forest catalog to date at these redshifts and increases by an order of magnitude the number of absorbers measured using the additional constraints provided by higher-order Lyman series transitions.

\begin{deluxetable}{cc}
\tablecaption{Median $\log(\NHI)$ uncertainty for absorber measurements}  
\tablewidth{0pt}
\tablehead{
\colhead{$\log(\NHI/ \rm cm^{-2})$} & \colhead{$\sigma_{\log(\NHI)}$} }
\startdata
   $  12.0  -  12.5 $ &   0.10  \\
   $  12.5  -  13.0 $ &   0.07  \\
   $  13.0  -  13.5 $ &   0.07  \\
   $  13.5  -  14.0 $ &   0.06  \\
   $  14.0  -  14.5 $ &   0.04  \\
   $  14.5  -  15.0 $ &   0.03  \\
   $  15.0  -  15.5 $ &   0.03  \\
   $  15.5  -  16.0 $ &   0.04  \\
   $  16.0  -  16.5 $ &   0.07  \\
   $  16.5  -  17.0 $ &   0.06  \\
   $  17.0  -  17.5 $ &   0.09 
 \enddata
     \label{NHI_error}
\end{deluxetable}

\section{\NHI\ Frequency Distribution}

\label{fN_text}

\begin{figure*}
\center
\includegraphics[width=0.7\textwidth]{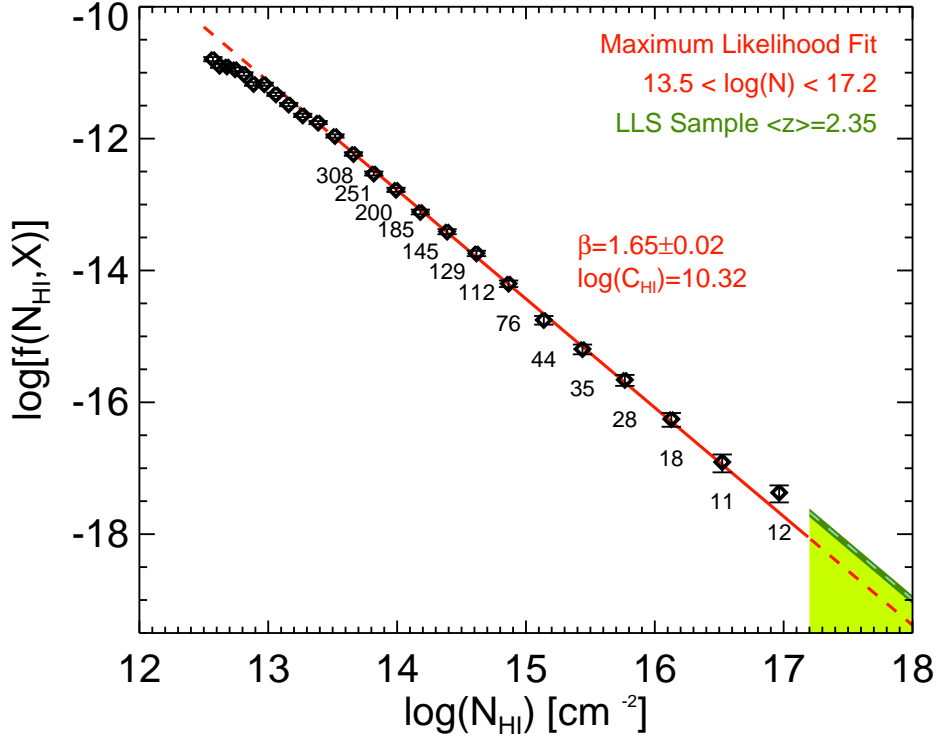}
\caption{\small The frequency distribution of absorbers from the KBSS \ion{H}{1} sample (black data points). For bins with $\NHI > 13.5$ \cm2, the number of absorbers per bin is displayed. The (red) line is a maximum likelihood power-law fit to the distribution of absorbers with $13.5 < \log(\NHI/ \rm cm^{-2}) < 17.2$. The (green) triangle represents the frequency of LLS calculated from other studies as discussed in \S \ref{LLS}. The dark green band at the top of the triangle represents the error in this estimation. Note this is a single power law fit to the data. Section \ref{broken_text} discusses broken-power law fits which better reproduce some aspects of the data, especially the incidence of LLSs.}
\label{FNX_LLS}
\end{figure*}

The differential column density distribution, first defined by \citet{car84}, is a useful way of parameterizing the frequency of absorbers as a function of \NHI. We consider the differential column density distribution per unit pathlength, $f(N_{\rm HI}, X)$, defined as 
\begin{equation}
f(N_{\rm HI}, X) dN_{\rm HI} dX = \frac{m}{\Delta N_{\rm HI} \Delta X} dN_{\rm HI}dX
\end{equation}
where $m$ is the observed number of absorbers with column densities in the range $\Delta \NHI$. $\Delta X$ is the comoving pathlength of the survey summed over the sightlines. This comoving pathlength was introduced by \citet{bah69} in part to understand if the statistics of absorbers were consistent with having been produced by galaxies. With this formalism, absorbers with a constant physical size and comoving number density will have constant $f(N,X)$.
The comoving pathlength of a single sightline, $\Delta X_i$, is defined as:
\begin{equation}
dX =  \frac{H_0}{H(z)}  (1+z)^2 dz
\end{equation}
\begin{equation}
\Delta X_i =  \int_{z_{min}}^{z_{max}} dX =\int_{z_{min}}^{z_{max}} \frac{(1+z)^2}{\sqrt{\Omega_\Lambda + \Omega_\textrm{\footnotesize{m}}(1+z)^3 }}  dz
\label{pathlength}
\end{equation}
 The total pathlength covered by this survey is $\Delta X = 26.9$ and the $\Delta X$ covered as a function of redshift summed over the 15 lines of sight is shown in Figure \ref{z_X}. The pathlength-weighted mean redshift of the sample is $\langle z \rangle = 2.37$. Measurements of $f(N,X)$ derived from the KBSS sample of absorbers with $12.5 < \log(\NHI/ \rm{cm}^{-2}) < 17.2$ are presented in Figure~\ref{FNX_LLS}. The data points are located at the mean $\log(\NHI)$ of the bin edges.




\subsection{Parameterizing the frequency distribution}

It has become standard to represent $f(\NHI,X)$ by a power law, or a series of broken power laws of the form:
\begin{equation}
f(N_{\rm HI}, X) dN_{\rm HI} dX = C_{\rm HI} N_{\rm HI}^{-\beta}dN_{\rm HI}dX.
\end{equation}
\citet{tyt87} first noted that the full column density distribution from $12<\log(\NHI/ \rm{cm}^{-2})<21$ was reasonably well approximated by a single power law with $\beta=1.5$. The detailed deviations from a single power law at various \NHI\ and the change in normalization and slope of $f(N_{\rm HI}, X) $ as a function of redshift have been considered by numerous authors \citep[e.g.,][]{sar89, pet93, dav01, kim02, pen04, jan06, leh07, pro10,rib11,ome12}.

The most notable deviations from a single power law appear at the ends of the distribution. The low-\NHI\ end appears to flatten, an effect generally attributed to a combination of incompleteness due to line blending and blanketing and a true physical turnover \citep{hu95,ell99, sch01,kim02}. The deviation from a single power-law for $\log(\NHI/ \rm{cm}^{-2}) \gtrsim 19$ is believed to be due to the transition of gas from optically thin to self-shielding regimes \citep{mur90,pet92,kat96,cor01,zhe02,alt11,mcQ11}. The most recent estimates of the frequency distribution of absorbers with $19 < \log(\NHI/ \rm{cm}^{-2}) \lesssim 20.3$ at $z \approx2$ are by \citet{ome07}; the frequency distribution of damped Lyman $\alpha$ systems with $\log(\NHI/ \rm{cm}^{-2}) > 20.3$ is measured to relatively high precision using the Sloan Digital Sky Survey (SDSS) QSOs by \citet{not09}. In this paper, we leverage the KBSS QSO absorption line data set to measure $f(N,X)$ for absorbers with $\log(\NHI/ \rm{cm}^{-2}) < 17.2$.

\subsection{Maximum Likelihood Method}

\label{mle}

\begin{figure*}
\center
\includegraphics[width=0.7\textwidth]{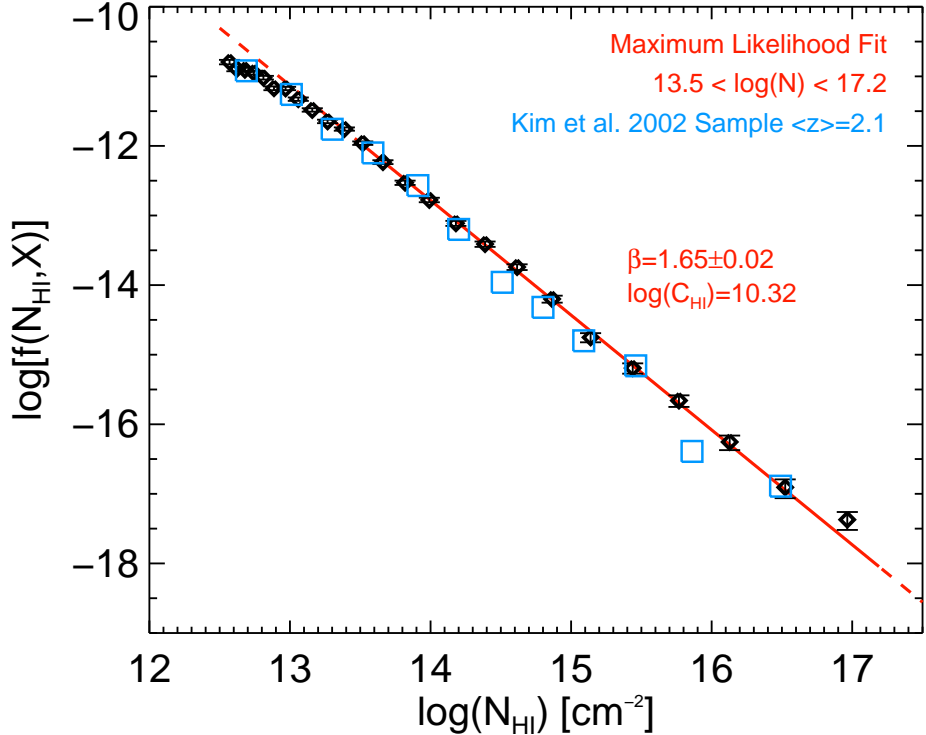}
\caption{\small The frequency distribution of absorbers from the KBSS \ion{H}{1} sample (black points) and the maximum likelihood fit to the data (red curve). For comparison, over plotted in (blue) squares are the data points from the $\langle z \rangle =2.1$ sample from \citet{kim02}. Because the analysis of the KBSS sample includes higher-order \ion{H}{1} transitions, we can measure the frequency of absorbers with $\log(\NHI/ \rm{cm}^{-2}) \gtrsim 14$ with much higher fidelity than has been previously possible. Notably, there is no evidence for a strong break in the power law distribution near $\log(\NHI/ \rm{cm}^{-2}) \approx 14$ as has been previously suggested.}
\label{fnx_kim02}
\end{figure*}

A statistically rigorous estimate of the power law exponent and normalization can be found via the Maximum Likelihood Method. The maximum likelihood estimator (MLE) for the power law exponent is:

\begin{equation}
\hat\beta=  1 + n \left[\sum\limits_{i=1}^n \ln \frac{N_i}{N_{\rm min}}\right]^{-1}
\end{equation}
where $n$ is the total number of absorbers in the sample and $N_{\rm min}$ in the column density of the lowest-column absorber \citep[see e.g.,][]{cla07}. The standard error on $\hat \beta$, derived from the width of the likelihood function, is:

\begin{equation}
\sigma_{\hat\beta}=  \frac{\hat\beta -1}{\sqrt{n}}
\end{equation}

Similarly, the estimator of the normalization, $C_{\rm HI}$, is

\begin{equation}
\hat C_{\rm HI} = \frac{n (1-\hat\beta)}{\left(N_{\rm max}^{(1-\hat\beta)} - N_{\rm min}^{(1-\hat\beta)}\right)\Sigma\Delta X}
\end{equation}
\citep{tyt87} where $\Sigma\Delta X$ is the sum of the pathlength over all 15 sightlines . 


\begin{deluxetable}{cccr}
\tablecaption{Maximum Likelihood Fits\tablenotemark{a}}  
\tablewidth{0pt}
\tablehead{
\colhead{$\log(N_{\rm HI, min})$} & \colhead{$\hat\beta$} & \colhead{$\sigma_{\hat\beta}$} & \colhead{$\log(\hat C_{\rm HI})$} }
\startdata
12.5 &  1.518 & 0.008 &   8.43 \\
12.6 &  1.544 & 0.008 &   8.80 \\
12.7 &  1.567 & 0.009 &   9.13 \\
12.8 &  1.585 & 0.010 &   9.38 \\
12.9 &  1.605 & 0.011 &   9.67 \\
13.0 &  1.616 & 0.012 &   9.82 \\
13.1 &  1.627 & 0.013 &   9.99 \\
13.2 &  1.636 & 0.014 &  10.11 \\
13.3 &  1.651 & 0.015 &  10.33 \\
13.4 &  1.658 & 0.016 &  10.44 \\
13.5 &  1.650 & 0.017 &  10.32 \\
13.6 &  1.636 & 0.018 &  10.11 \\
13.7 &  1.639 & 0.020 &  10.15 \\
13.8 &  1.645 & 0.021 &  10.25 \\
13.9 &  1.649 & 0.023 &  10.31 \\
14.0 &  1.654 & 0.025 &  10.39 
 \enddata
 \tablenotetext{a}{Maximum likelihood fits to the data with $\log(N_{\rm HI ,min})$ $\le \log(\NHI/ \rm{cm}^{-2}) \le 17.2$.}
     \label{mle_tab}
\end{deluxetable}

In Figure \ref{FNX_LLS} the MLE fit to the data is shown for absorbers with $13.5 < \log(\NHI/ \rm{cm}^{-2}) < 17.2$.  Note that a single power law reproduces the distribution with reasonable fidelity. No strong breaks in the distribution are evident. 

The choice of a minimum \NHI$=13.5$ used in the fit is motivated as follows. As shown in Figure \ref{FNX_LLS}, $f(\NHI,X)$  for $\log(\NHI/ \rm{cm}^{-2}) \lesssim 13.5$ appears to flatten. As mentioned above, this flattening has previously been observed \citep{hu95,ell99, kim02} and is generally attributed to a combination of incompleteness in the sample caused by line blending and a true physical turnover in $f(\NHI,X)$. Here we do not attempt to diagnose the cause of the low-\NHI\ turnover, but simply note that above $\log(\NHI/ \rm{cm}^{-2}) \ge 13.5$, $f(N,X)$ appears to be well fit by a single power law. More quantitatively, one can measure the change in power law slope introduced by varying the minimum value of \NHI\ included in the fit as shown in Table \ref{mle_tab}.  Note that a choice of $\log(\NHI/ \rm{cm}^{-2}) \gtrsim 13.3$ produces insignificant change in the maximum likelihood estimate of the slope, $\hat\beta$.

Prior to the present work, the largest existing sample of Ly$\alpha$ forest absorbers at these redshifts was described by \citet{kim02}. Their sample included 2130 \ion{H}{1} absorbers with $12.5 < \log(\NHI/ \rm{cm}^{-2})<17$ observed along 9 lines of sight. The most marked improvement over the previous measurements provided by our study is the use of higher-order transitions of \ion{H}{1} which were generally not included in previous work. These allow for more precise measurements of \NHI\ above the saturation point of Ly$\alpha$, $\NHI \approx 10^{14-14.5}$ \cm2. Above $\log(\NHI/ \rm{cm}^{-2}) \gtrsim 15.5$ where the Ly$\beta$ transition saturates, the sample presented herein may suffer a small degree of incompleteness at $z\lesssim2.4$ as discussed in Appendix \ref{App_fitting}. 

In Figure \ref{fnx_kim02}, the KBSS data (black diamonds) are compared with the measurements of \citet{kim02} at $\langle z \rangle = 2.1$ (blue squares). Notably, at $\log(\NHI/ \rm{cm}^{-2}) < 14$ there is good agreement between the two data sets, but for $\log(\NHI/ \rm{cm}^{-2}) \gtrsim 14$, the \citet{kim02} sample exhibits significant scatter not present in the KBSS measurements, with a systematic tendency to under-estimate the number of intermediate-\NHI\ ($14 \lesssim \log(\NHI/ \rm{cm}^{-2}) \lesssim 17)$ absorbers compared to our measurements.  Potential incompleteness found in the lower-$z$ portion of the catalog presented herein (see Appendix \ref{App_fitting}) may also affect the \citet{kim02} catalog for absorbers with $\log(\NHI/ \rm{cm}^{-2})\gtrsim14$ and may explain why their measurements are systematically lower than those presented here.

\subsection{Lyman Limit Systems}
\label{LLS}

In this section, we quantify the incidence of absorbers with \NHI $> 10^{17.2}$ cm$^{-2}$, Lyman Limit Systems (LLSs) which are optically thick to hydrogen-ionizing photons.  Because LLSs are relatively rare, and because their \NHI\ are challenging to determine accurately via line fitting, we use samples reported in the literature to measure the frequency of optically thick absorbers at the same redshifts as the KBSS sample. 

We estimated the incidence of LLSs with $\log(\NHI/ \rm{cm}^{-2}) > 17.2$ at $\langle z \rangle \sim 2.4$ using data from two surveys that nicely bracket the relevant range of redshifts for our IGM sample. We evaluated the LLSs over a somewhat wider redshift range than $2 \simlt z \simlt 2.8$ (but having the same mean redshift) in order to mitigate the statistics of small numbers.   Over the range $1.65 < z < 2.55$, we used data from \citet{rib11} from their survey compilation of archival HST spectroscopy; results for $2.70 < z < 2.95$ are based on a subset of the ground-based surveys of \citet{sar89} and \citet{ste95}. The sub-samples are roughly equivalent in their statistical significance (comparable total pathlength and number of LLSs), with path-weighted mean redshifts of $\langle z \rangle = 2.08$ and $\langle z \rangle = 2.84$, respectively. The path-weighted mean redshift for the full sample is $\langle z_{\rm LLS} \rangle = 2.41$ based on 46 LLSs with $\log(\NHI/ \rm{cm}^{-2}) > 17.2$ in a total redshift path of $\Delta z = 25.95$ ($\Delta X = 88.31$).  These LLS samples provide a maximum likelihood estimate of the number of LLS per unit pathlength, $dn_{\rm LLS}/dX = 0.52 \pm 0.08$.

A recent study of LLS near the redshift range of interest ($2.0 < z < 2.6$)  using uniform HST data is described in \citet{ome12}. They measure an integral constraint on the number of LLS absorbers with $\tau_{912}>1, \log(\NHI/ \rm{cm}^{-2}) > 17.2$ and $\tau_{912}>2, \log(\NHI/ \rm{cm}^{-2})>17.5$, where $\tau_{912}$ is the optical depth at the Lyman Limit, 912\AA, in the absorber's rest frame. These authors found $dn_{\tau>1}/dX =  0.48^{+0.09}_{-0.12}$, in excellent agreement with our compilation described above ($dn_{\rm LLS}/dX = 0.52 \pm 0.08$). For more optically thick systems, \citet{ome12} found $dn_{\tau>2}/dX = 0.28^{+0.06}_{-0.06}$. The consistency of these measurements with the lower-\NHI\ frequency distribution will be discussed further in \S \ref{LyC}.

\begin{figure*}
\center
\includegraphics[width=0.65\textwidth]{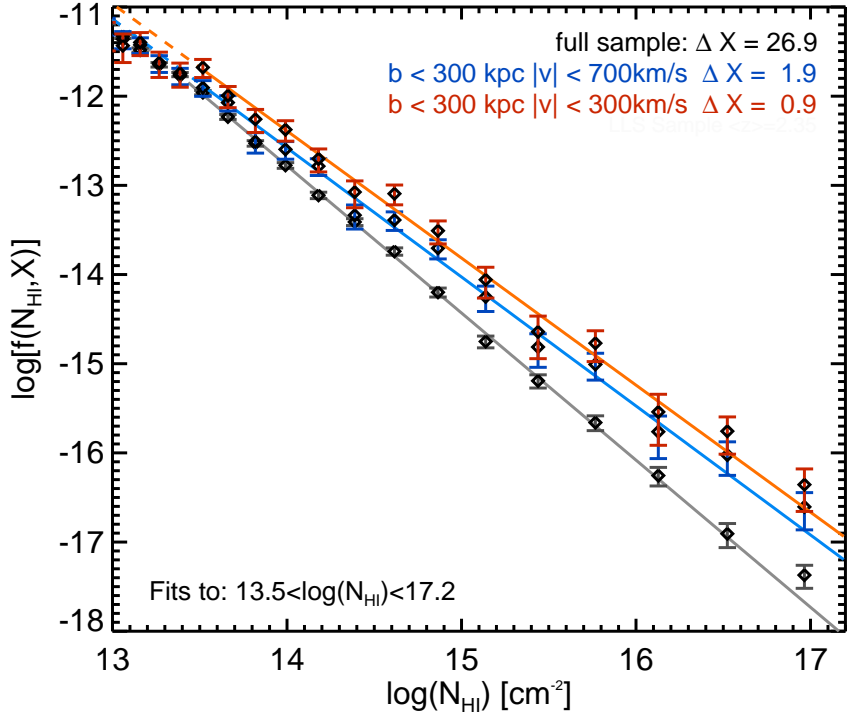}
\caption{\small Same as Figure \ref{FNX_LLS} but also considering the distribution of absorbers close to galaxies. The IGM distribution is shown in grey. Red points show the frequency of absorbers of a given \NHI\ within 300 \kms\ and 300 pkpc of a galaxy in our spectroscopic sample. Blue points show the frequency of those within 700 \kms\ and 300 pkpc. The curves are the maximum likelihood fits to the distributions over the range $13.5 < \log(\NHI/ \rm{cm}^{-2}) < 17.2$. The parameters of the fit are given in Table \ref{mle_tab} along with the opacities calculated from the data and the opacities implied by the fit. Notably this single power law fit to the data does not well reproduce the opacities calculated from the data. Section \ref{broken_text} discusses broken-power law fits which better reproduce these opacities.}
\label{fNX_CGM_mle}
\end{figure*}


Unfortunately, surveys that rely on detecting breaks in the spectrum of QSOs at the redshift of the LLS can rarely measure \NHI\ accurately, especially for absorbers with $\tau_{912}>2$. As such, a measurement of the power law index of $f(N,X)$ in the LLS regime typically relies on the combination of integral constraints ($dn/dX$) for the LLS, subDLA, and DLA regimes, as well as the measured values of $f(N,X)$ for Ly$\alpha$ forest absorbers. As we now have much more statistically robust measurements of $\beta$ for Ly$\alpha$ forest absorbers with $ \log(\NHI/ \rm{cm}^{-2}) \lesssim 17.2$, we re-examine the power law fit to $f(N,X)$ in the LLS regime.  

Here, we test the consistency of the measured incidence of LLSs with the extrapolated slope inferred from the MLE method for absorbers with $\log(\NHI/\rm cm^{-2}) < 17.2$. First, we note that LLS surveys are sensitive to all absorbers with $ \log(\NHI/ \rm{cm}^{-2}) >  17.2$. Observations of DLAs \citep[e.g.,][]{not09} show that there is a strong turnover in the frequency of DLAs at $ \log(\NHI/ \rm{cm}^{-2}) \gtrsim 21$, suggesting there are very few absorbers with $ \log(\NHI/ \rm{cm}^{-2}) \gtrsim 22$. Here, we assume that $dn_{\rm LLS}/dX$ is sensitive to absorbers with  $ 17.2 < \log(\NHI/ \rm{cm}^{-2}) >  22.0$. 

Next, we consider the incidence of LLSs that would be found if the MLE power law fit constrained by our data at $ \log(\NHI/ \rm{cm}^{-2}) < 17.2$ held for the entire LLS regime. We calculate the incidence of LLSs given the fit parameters as follows:
\begin{eqnarray}
dn_{\rm LLS}/dX &=& \int_{N_{\rm min}}^{N_{\rm max}} f(N_{\rm HI}, X) d\NHI \\
dn_{\rm LLS, fit}/dX &=& \int_{N_{\rm min}}^{N_{\rm max}} C_{\rm HI} N_{\rm HI}^{-\beta} d\NHI.
\label{LLS_eqn}
\end{eqnarray}
where $\log(N_{\rm min}/ \rm{cm}^{-2})=17.2$ and $\log(N_{\rm max}/ \rm{cm}^{-2})=22.0$.
We find that this extrapolation to $\log(\NHI/ \rm{cm}^{-2}) > 17.2$ underestimates the number of LLSs significantly giving $dn_{\rm LLS,fit}/dX = 0.21$ (compared to $dn_{\rm LLS,data}/dX = 0.52 \pm 0.08$).

The green triangle in Figure \ref{FNX_LLS} represents a similar exercise graphically on the $f(N,X)$ plot. If we assume that the MLE determination of $\beta$ for the forest holds in the LLS regime, we can calculate the value of $C_{\rm HI, LLS}$ implied by the measured $dn_{\rm LLS}/dX$ by inverting equation \ref{LLS_eqn}. Then the values of $f(N,X)$ suggested by the data would be given by:
\begin{equation}
f(N_{\rm p},X) = C_{\rm HI, LLS} N_{\rm p}^{-\beta}.
\end{equation}
for all values of $N_{\rm p}$.
The height of the green triangle thus represents the values of $f(N,X)$ if $\beta_{\rm MLE}$ holds of the LLS regime. The dark green band at the top of the triangle is an estimate in the uncertainty of this normalization due to the uncertainty in $dn_{\rm LLS}/dX$. Again, we can see clearly that the MLE fit to the frequency distribution (red dotted line) significantly under-predicts the measured LLS incidence.  

We have seen that a single power law well approximates the intermediate-\NHI\ absorbers; however, its extrapolation to $\log(\NHI/ \rm{cm}^{-2}) > 17.2$ underestimates the number of LLSs significantly giving $dn_{\rm LLS,fit}/dX = 0.21$. In \S \ref{LyC}, we discuss other methods for parameterizing the distribution which better reproduce the high-\NHI\ end of the frequency distribution.

\section{Frequency Distributions with Respect to the Positions of Galaxies}

\label{gal_fN}

\citet{gcr12} considered the distribution of \ion{H}{1} with respect to the positions of star-forming galaxies with $2.0 < z < 2.8$ using the absorption line data presented here. The study explored the kinematic and geometric structure of the CGM, concluding that the majority of the strong excess absorption is found within 300 physical kpc (pkpc) of galaxies in the transverse direction and within 300 \kms\ of the systemic redshift of galaxies; however, a tail of excess \ion{H}{1} absorbers continues to 2 pMpc and 700 \kms. 

The KBSS spectroscopic galaxy sample consist of UV-color selected galaxies with UV luminosities $0.25 < L/L^* < 3.0$ \citep{red09} and a typical bolometric luminosity of $\sim2.5 \times 10^{11} L_\odot$ \citep{red08, red12}. They have star formation rates of $\sim30$ \Msun\ yr$^{-1}$ \citep{erb06c} and halo masses of $\sim10^{12}$ \Msun\ \citep[][O. Rakic et al. 2012 in preparation]{ade05b,con08,tra12}.

Using the same KBSS galaxy sample as in \citet{gcr12}, we measure the frequency distribution of \ion{H}{1} absorbers within the CGM as follows. First we locate those galaxies in our spectroscopic sample lying within 300 pkpc of the line of sight to the QSO and within the redshift range used for the Ly$\alpha$ forest decompositions (as listed in Table \ref{field}). Notably, we have no galaxies at impact parameters $< 50$ pkpc, and so the CGM distributions apply to the range $50 - 300$ pkpc. 
We then consider the column density distribution of absorbers within $\pm 300$ \kms\ and $\pm 700$ \kms\ of the redshifts of these galaxies. We note that all absorbers included in the 300 \kms\ CGM are also included in the 700 \kms\ CGM.

The resulting frequency distributions for the  300 \kms\ and 700 \kms\ CGM are shown in Figure \ref{fNX_CGM_mle} in red and blue respectively. As found by \citet{gcr12}, the frequency of all absorbers with $\log(\NHI/ \rm{cm}^{-2}) > 13.5$ is higher within the CGM of galaxies than at random places in the IGM. 

The maximum likelihood estimates of the power law approximation to $f(\NHI, X)$ within the CGM are listed in Table \ref{mle_tab} and overplotted in Figure \ref{fNX_CGM_mle}. Notably, the power law slope of the CGM is significantly shallower than that of the IGM, in the sense that there are relatively more high-\NHI\ systems than low-\NHI\ absorbers in the CGM. This result is qualitatively equivalent to results presented by \citet{gcr12} who found a strong correlation between \NHI\ and the fraction of absorbers found within the CGM of galaxies (their Figure 30). Further, the frequency of absorbers is higher within the 300 \kms\ CGM than within 700 \kms, as expected.\footnote{Since the majority of the excess absorption is found within 300 \kms\ of galaxies, the 700 \kms\ CGM is diluted by regions with \ion{H}{1} properties more similar to the general IGM. As such, the difference between the 300 and 700 \kms\ CGM is mostly a result of the difference in pathlength considered. The pathlength for the 300 \kms\ CGM sample is $\Delta X = 0.9$ while the 700 \kms\ sample has twice the pathlength, $\Delta X = 1.9$ - and since \fnx\ is inversely proportional to the pathlength, the frequency of absorbers in the 700 \kms\ CGM is approximately a factor of 2 smaller. }

\begin{deluxetable*}{lcccr}
\tablecaption{MLE Power Law fits to $f(\NHI,X)$}  
\tablewidth{0pt}
\tablehead{ 
\colhead{Sample} & \colhead{$\beta$} & \colhead{$\log(C_{\rm HI})$} & \colhead{$\kappa_{\rm fit}$ [pMpc$^{-1}$]} & \colhead{$\kappa_{\rm data} \tablenotemark{a}$  [pMpc$^{-1}$]}}\startdata

IGM & $1.650 \pm 0.017$ & 10.322 & 0.0029 & $0.0039^{+0.0005}_{-0.0005}$\\\\
CGM 700km s$^{-1}$ 300 pkpc  &  $1.447 \pm 0.033$ &  7.672 & 0.013 &  $0.017^{+0.004}_{-0.004}$\\\\
CGM 300km s$^{-1}$ 300 pkpc &  $1.425 \pm 0.037$ &  7.562 & 0.022 &  $0.031^{+0.009}_{-0.009}$\\
 \enddata
 \tablenotetext{a}{Opacity calculated from the the data using equation \ref{kappa_data} for $13.5 \le \log(\NHI/ \rm{cm}^{-2}) \le 17.2$.}
     \label{mle_tab}
\end{deluxetable*}

\begin{figure*}
\centerline{
\includegraphics[width=0.45\textwidth]{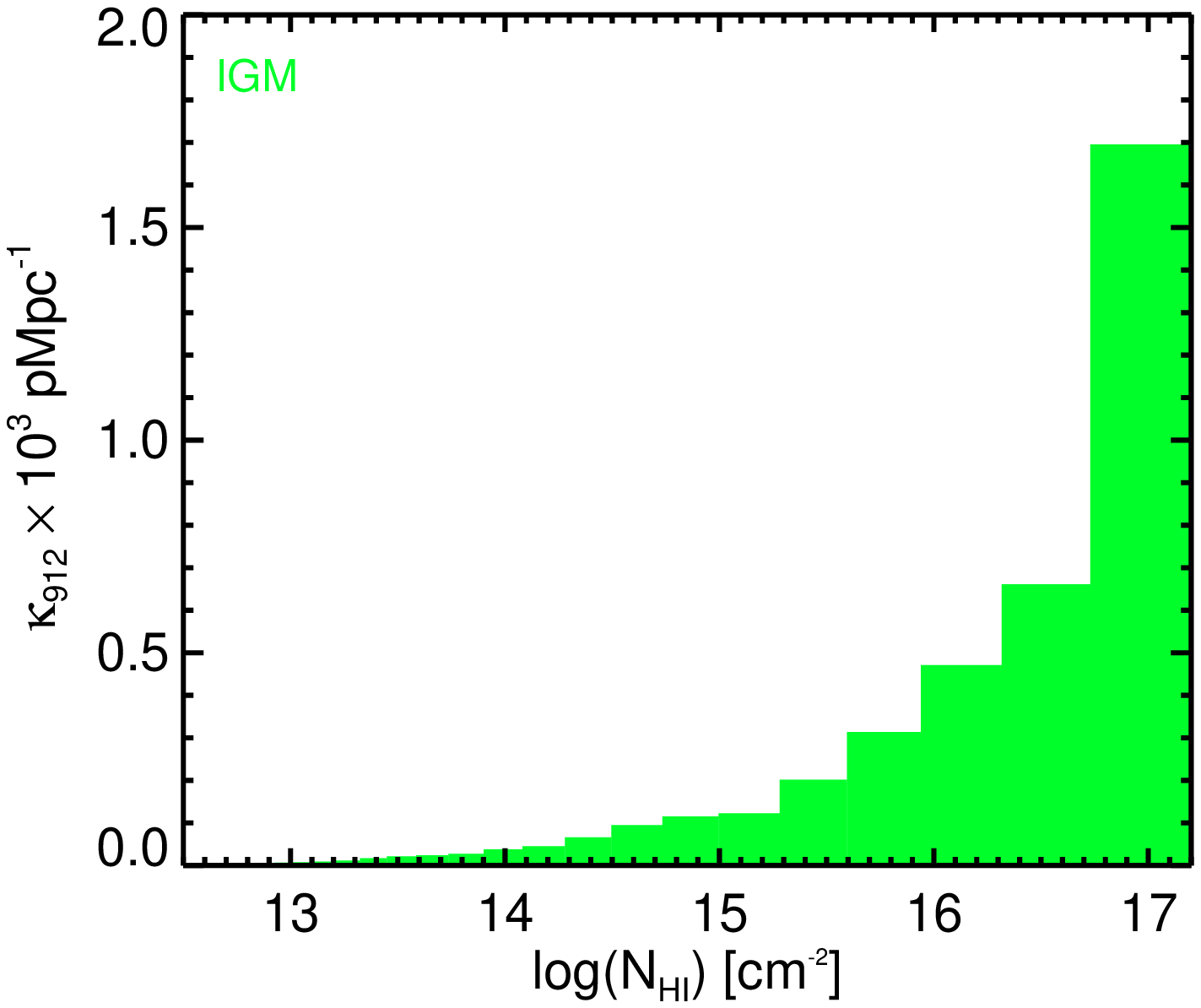} \includegraphics[width=0.45\textwidth]{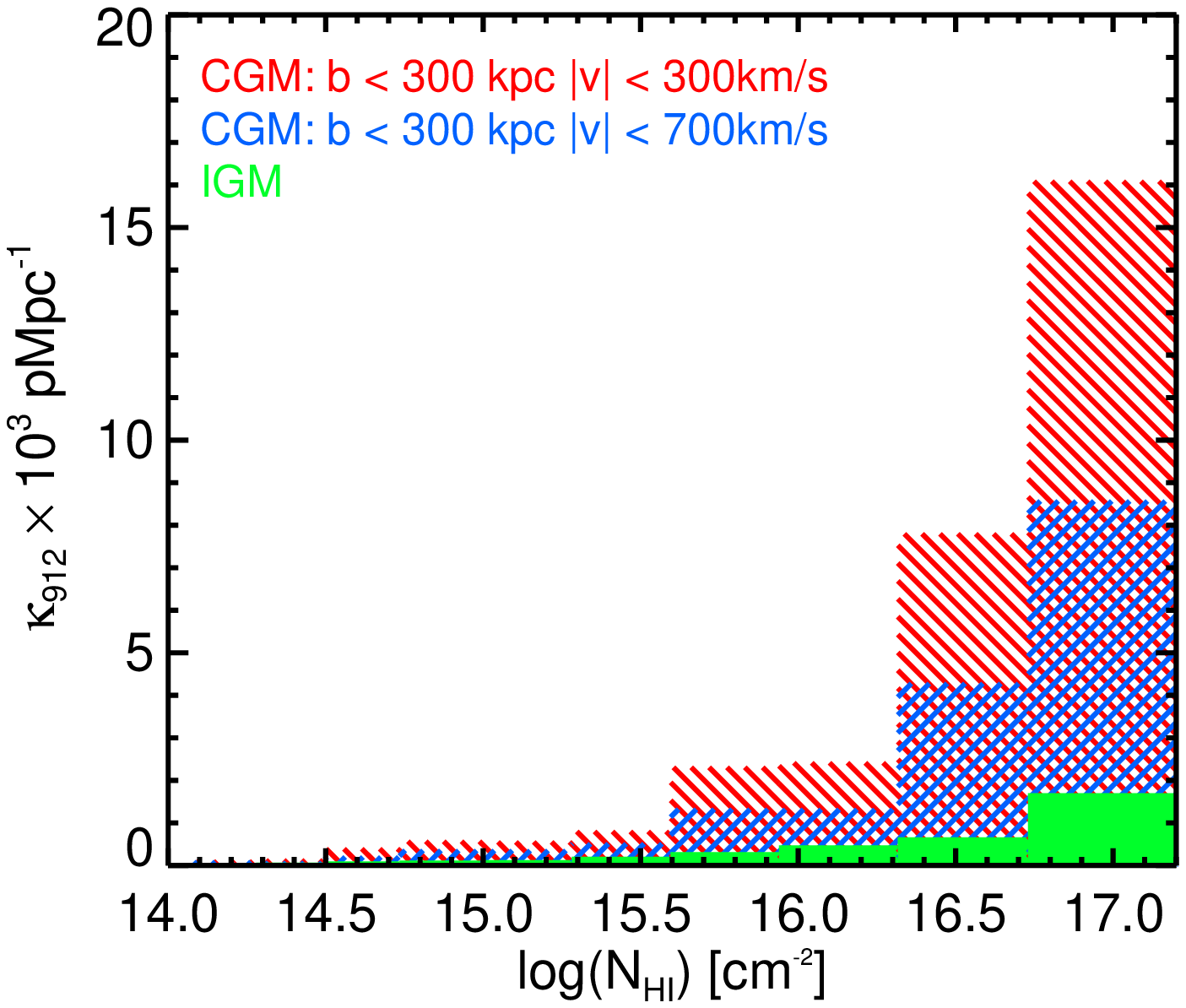}}
\caption{\small Contributions to the opacity from a given bin in \NHI, measured directly from the data using equation \ref{kappa_data}. \textit{Left:} The IGM opacity per bin in \NHI. \textit{Right:} The opacity per bin in \NHI\ within the IGM (green), the 700 \kms\ CGM (blue) and the 300 \kms\ CGM (red). Note that the scale of the left-hand panel is 10\% of the scale of the right-hand panel. The opacity (optical depth per unit physical distance) of the CGM is an order of magnitude larger than the average opacity in the IGM. }
\label{kappa_data_hist}
\end{figure*}

The characteristic \NHI\ of the CGM is akin to the max(\NHI) statistic presented in the left-hand panel of Figure 9 of \citet{gcr12}. These authors typically found $\log(\NHI/ \rm{cm}^{-2}) \gtrsim 14.5-15$ within 300 pkpc of a galaxy. \citet{mis07} considered the frequency distribution of absorbers surrounding \ion{H}{1} systems with $\log(\NHI/ \rm{cm}^{-2}) > 15$, fitting to the absorber distribution within $\pm 1000$ \kms of the redshifts of these tracer absorbers. Thus, the  \citet{mis07} sample traces a volume similar to that of the CGM. For their S2a sample which was selected to lie at least 5000 \kms\ lower than the redshifts of the QSOs, \citet{mis07} find a power-law slope of $\beta = 1.390 \pm 0.027$, in reasonable agreement with the power-law exponents for our CGM measurements. The normalization, C$_{\rm HI} = 7.262 \pm 0.431$, is lower than our CGM results, consistent with the expectation that the \citet{mis07} sample includes a larger fraction of ``general'' IGM compared to CGM.

\section{The Lyman Continuum Opacity}
\label{LyC}

One of the principal uses of the column density distribution presented above is to estimate the transmissivity of the IGM to hydrogen-ionizing photons, hereafter referred to as Lyman Continuum (LyC) photons.  Understanding LyC emission and its transmission through the IGM is an important step in exploring the process of reionization at high-redshift, as well as estimating the average ionization level of the IGM and the metagalactic ionizing background as a function of redshift.


In order to understand and describe the process of transmission of LyC photons through the IGM, one must first quantify the sources of opacity in the IGM. Most notably, it is important to quantify the fractional contribution of absorbers of various \NHI\ to the full opacity of the forest. To do this, we begin with a simplified model of LyC emission and absorption. 

The attenuation of LyC photons by an absorber with \NHI$=N_i$ is given by
$e^{-N_i \sigma}$
where $\sigma$ is the photoionization cross section of hydrogen. In practice, $\sigma$ depends sensitively on the energy of the photons considered:
\begin{equation}
\sigma(\nu) \approx \sigma_{\rm LL} \left(  \frac{\nu}{\nu_{\rm LL}} \right)^{-3}
\end{equation}
where $\sigma_{\rm LL} = 6.35 \times 10^{-18}$ cm$^2$ is the hydrogen ionization cross section to photons at the Lyman Limit (912 \AA) with energies of 1 Ry (13.6 eV), and $v_{\rm LL}$ is the frequency of such photons. For the moment, we will employ the simplifying assumption that all LyC photons have energies of 1 Ry so that the frequency dependence of $\sigma$ is removed. 

In addition, for the time being, we calculate the opacity assuming that the universe is static.\footnote{For the remainder of Section \ref{LyC} we use the redshift distribution of the sample to infer the physical distance over which absorbers are distributed. This is the only way in which redshifts enter into the calculation.} In practice, the expansion of the universe results in the redshifting of ionizing photons to non-ionizing energies; however since we are only considering 1 Ry photons, we will also temporarily ignore redshifting.  In \S \ref{mcText}, we perform a more rigorous calculation to measure the true mean free path of LyC photons through the IGM and CGM at z=2.4 including photons of all ionizing energies and including redshifting. The analytic calculation presented in this section is provided to build intuition regarding the sources of opacity in the universe, and also to quantify the opacity from absorbers with $\log(\NHI/ \rm{cm}^{-2}) < 17.2$ inferred directly from the data.

Using the above approximations, we calculate the opacity, $\kappa$, of the IGM and CGM to LyC photons and the contribution to this opacity from absorbers of various \NHI.  The opacity is defined as the change in optical depth per unit proper distance:
\begin{equation}
\kappa = \frac{d\tau}{dr}
\label{eqn_dt_dr}
\end{equation}
where 
\begin{equation}
dr =\frac{1}{1+z} \frac{c}{H(z)}dz
\end{equation}
and $c$ is the speed of light.

Given an analytic fit to the frequency distribution per unit physical distance $f(N_{\rm HI}, r)$, the opacity due to Poisson-distributed absorbers with $N_{\rm min} \le \NHI \le N_{\rm max}$ to photons of energy 13.6 eV is:
\begin{equation}
\kappa_{\rm fit} = \int_{N_{\rm min}}^{N_{\rm max}} f(N_{\rm HI}, r) \left(1-e^{-N_{\rm HI} \sigma_{912}}\right)dN_{\rm HI}
\label{kappa_fit}
\end{equation}
\citep{par80}.

\begin{figure*}
\center
\includegraphics[width=0.7\textwidth]{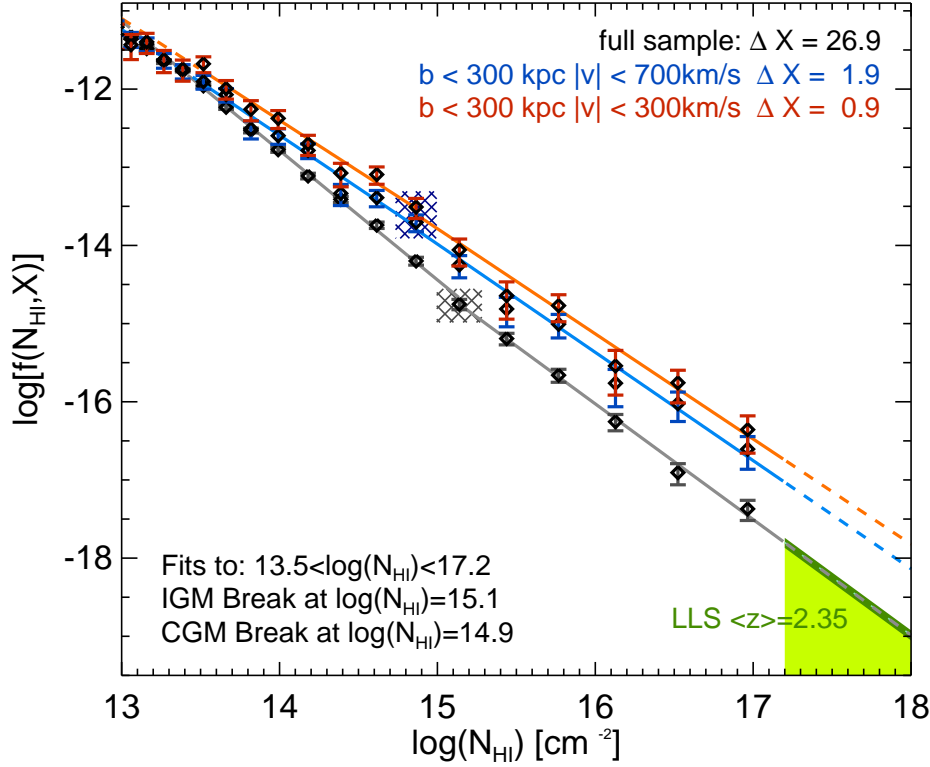}
\caption{\small Broken power law least-squares fits to the binned IGM (grey) and CGM (blue and red) data. The slopes and normalizations for each data set (IGM or CGM) are allowed to vary independently; however,  both ranges include the data point centered at the break (shaded regions). Note that the broken power-law fit is discontinuous at the break location. Including one data point in both ranges minimizes the discontinuity associated with the break and produces a better fit to the data. The hatched box shows the range of \NHI\ included in that point and thus in both fits. The green triangle represents the measured incidence of LLSs as discussed in \S \ref{LLS}.  The parameters of the fit are given in Table \ref{broken_tab} along with the opacity calculated from the data and the opacity produced by the fit to the distribution. Note that the opacity is reproduced to much higher fidelity using the broken power law fit.}
\label{broken_CGM}
\end{figure*}

Similarly, the opacity can be calculated directly from the data by summing over those absorbers \{i\} with $N_{\rm min}<\NHI<N_{\rm max}$:
\begin{equation}
\kappa_{\rm data} = \frac{ \sum_{i} \left(1-e^{-N_{\rm HI, i} \sigma_{912}}\right)}{\Delta r}
\label{kappa_data}
\end{equation}
where $\Delta r$ is the sum of the proper distances probed by each sightline.

The uncertainty in the measurement of $\kappa_{\rm data}$ can be calculated using the bootstrap resampling method in which many sets of absorbers are drawn with replacement and $\kappa_{\rm data}$ is recalculated for each set. One can then estimate the probability distribution of $\kappa_{\rm data}$ by selecting the 68\% confidence range. Values of $\kappa_{\rm data}$ for the IGM and CGM for $13.5 < \log(\NHI/ \rm{cm}^{-2}) < 17.2$ is listed in the last column of Table \ref{mle_tab}. \textit{We emphasize that this measurement of the LyC opacity from absorbers with $ \log(\NHI/ \rm cm^{-2})<17$  is one of the principal results of this work. }


The left hand panel of Figure \ref{kappa_data_hist} shows the opacity produced by various bins of $\log(\NHI/ \rm{cm}^{-2}) < 17.2$ within the IGM calculated directly from the data using equation \ref{kappa_data}. These values are compared to the opacity found within the CGM in the right-hand panel of Figure \ref{kappa_data_hist}. Notably, the opacity of the CGM of a galaxy (300 \kms, 300 pkpc) is an order of magnitude larger than that of the average IGM.

\begin{deluxetable*}{lcccccccr}
\tablecaption{Broken Power Law fits to $f(\NHI,X)$}  
\tablewidth{0pt}
\tablehead{
\colhead{Sample} & \colhead{$\beta_{\rm low-\NHI}$} & \colhead{$\log(C_{\rm low-\NHI})$} & \colhead{break \NHI} & \colhead{$\beta_{\rm high-\NHI}$} & \colhead{$\log(C_{\rm high-\NHI})$} & \colhead{$\kappa_{\rm fit}$ [pMpc$^{-1}$]} & \colhead{$\kappa_{\rm data}$\tablenotemark{a} [pMpc$^{-1}$]}}
\startdata

IGM: & 1.656$\pm$ 0.030 & 10.398 & 15.14 & 1.479$\pm$ 0.042 &  7.643 & 0.0038 & $0.0039^{+0.0005}_{-0.0005}$\\\\
CGM 700km s$^{-1} $ 300 pkpc &  1.353$\pm$ 0.088 &  6.354 & 14.87 & 1.385$\pm$ 0.084 &  6.789 & 0.017 & $0.017^{+0.004}_{-0.004}$\\\\
CGM 300km s$^{-1} $ 300 pkpc &  1.302$\pm$ 0.084 &  5.823 & 14.87 & 1.345$\pm$ 0.089 &  6.394 & 0.031 & $0.031^{+0.009}_{-0.009}$\\

 \enddata
 \tablenotetext{a}{Opacity calculated from the the data using equation \ref{kappa_data} for $13.5 \le \log(\NHI/ \rm{cm}^{-2}) \le 17.2$.}
     \label{broken_tab}
\end{deluxetable*}

Comparing the opacities calculated using the single power law MLE fit with those calculated directly from the data for absorbers with $\log(\NHI/ \rm{cm}^{-2}) < 17.2$ (Table \ref{mle_tab}), one notes a statistically significant discrepancy in the sense that the MLE power law predicts far less opacity than is measured directly from the data. Similarly, the number of LLSs (which also contribute significantly to the LyC opacity) is under-predicted by the MLE fit. In Section \ref{broken_text}, we consider other parameterizations of the frequency distribution that better match $dn_{\rm LLS}/dX$ as well as the opacity from intermediate-\NHI\ systems [$15 \lesssim \log(\NHI/ \rm{cm}^{-2}) \lesssim 17$] calculated from the data.

\subsection{Broken Power Law Parameterizations}
\label{broken_text}

While the MLE power law fits presented in \S \ref{mle} and \S \ref{gal_fN} provide a reasonable description of the overall column density distribution, because there are substantially more absorbers at low-\NHI\ the majority of the statistical power for the fit is derived from the low-\NHI\ absorbers. To estimate the LyC opacity of the intergalactic and circumgalactic gas, the absorbers of relevance are instead the much rarer, high-\NHI\ systems, particularly those near $\log(\NHI/ \rm{cm}^{-2}) \approx 17$.

 \begin{deluxetable*}{lcccc}
\tablecaption{Effect of uncertainty in $\beta$ on $\kappa$ and $dn_{\rm LLS}/dX$\tablenotemark{a} }  
\tablewidth{0pt}
\tablehead{
\colhead{Sample} & \colhead{$\beta_{\rm high-\NHI}$} & \colhead{$\log(C_{\rm high-\NHI})$\tablenotemark{b}} & \colhead{$\kappa$ [pMpc$^{-1}$]\tablenotemark{c}} & $dn_{\rm LLS}/dX$\tablenotemark{d}}
\startdata
Data\tablenotemark{e}    &   \nodata  & \nodata & $0.0039^{+0.0005}_{-0.0005}$ & $0.52 \pm 0.08$\\\\
IGM Best Fit & 1.479 &  7.643 & 0.0038 & 0.519 \\
IGM Shallow Fit\tablenotemark{f}  &   1.437 &      6.984   &   0.0040 &   0.656 \\
IGM Steep Fit\tablenotemark{g}  & 1.521 &     8.301    &   0.0035  &  0.413
 \enddata
    \label{beta_marg}
    \tablenotetext{a}{The effect of varying $\beta_{\rm high-\NHI}$ within its 1-$\sigma$ uncertainty.}
    \tablenotetext{b}{The best fit normalization of $f(N,X)$ given the stated value of $\beta$. I.e., the covariance of $\log(C_{\rm high-\NHI})$ with $\beta_{\rm high-\NHI}$. }
    \tablenotetext{c}{The opacity from absorbers with $\log(\NHI/\rm cm^{-2}) < 17.2$.}
    \tablenotetext{d}{The incidence of LLSs}
    \tablenotetext{e}{Values computed directly from the data.}
    \tablenotetext{f}{The fit for the perturbed value of $\beta_{\rm high-\NHI}$ such that the distribution is more shallow.}
    \tablenotetext{g}{The fit for the perturbed value of $\beta_{\rm high-\NHI}$ such that the distribution is steeper.}
\end{deluxetable*}

In order to reproduce simultaneously the opacity from both absorbers with $12.5 < \log(\NHI/ \rm{cm}^{-2})<17.2$ as measured in our sample, as well as the number of LLSs, we adopt a different fitting method. Here we consider a least-squares fit to the binned\footnote{ We have verified that the fitted parameters are insensitive to the exact choice of binning.} 
 data. Because the opacity is a strong function of \NHI, it is most important that the number of absorbers with $\log(\NHI/\rm cm^{-2}) \approx 17.2$ is accurately reproduced. Therefore, 
 to provide further flexibility, we allow for a discontinuous break in the power law parameterization. This allows for an independent fit to the high-\NHI\ absorbers [$\log(\NHI/ \rm cm^{-2}) \gtrsim 15$] which are the main source of opacity. 
 Although no strong break is evident, any parameterization that does not include a break as well as all parameterizations with a continuous functional form at the position of the break cannot reproduce the opacity of Ly$\alpha$ forest absorbers as measured by the data. 

We fit two power laws to the binned data. The bin that includes the break is included in the fit to the low and intermediate-\NHI\ groups to minimize the discontinuity between the two power laws. We allow the break to vary in \NHI\ and select the break location to be that which best reproduces the opacity from absorbers with $\log(\NHI/ \rm{cm}^{-2}) < 17.2$ and also the number of LLS in the case of the IGM sample. The break location for the two CGM distributions is forced to match. The best fits are presented in Figure \ref{broken_CGM} and Table \ref{broken_tab} along with their implied opacities, $\kappa_{\rm fit}$.  The bin containing the break is indicated by the hatched box in Figure \ref{broken_CGM}. Notably, the location of the break ($\NHI \approx 10^{15}$ \cm2) is  similar to the \NHI\ at which absorbers were found to significantly correlate with the positions of galaxies \citep{gcr12}.

This fitting technique reproduces the opacity from intermediate-\NHI\ absorbers measured from the data with high fidelity. Additionally, the IGM fit predicts a $dn_{\rm LLS,fit}/dX = 0.519$ for $17.2 < \log(\NHI/ \rm{cm}^{-2}) < 22.0$ in excellent agreement with  $dn_{\rm LLS,data}/dX =0.52 \pm 0.08$ determined from the compilation of LLSs from the literature (Section \ref{LLS}).\footnote{We remind the reader that the single power law MLE determination of the fit predicted a value of $dn_{\rm LLS,fit}/dX =0.21$.}  As in section \ref{LLS}, the green triangle in Figure \ref{broken_CGM} represents the values of $f(N,X)$ within the LLS regime implied by the measurement of $dn_{\rm LLS,data}/dX$ \textit{assuming a power law index,} $\beta=\beta_{\rm high-\NHI}$. The agreement between the height on the green triangle and the extrapolation of the power law fit (grey line) suggests that the broken power law parameterization well reproduces the LLS gross statistics.

Next we consider the uncertainty in the fitted parameters. The majority of the uncertainty in the normalization of the frequency distribution ($C_{\rm \NHI}$) is due to its covariance with the power law index, $\beta$. In Table \ref{beta_marg} we list the best fit determination of $\beta_{\rm high-\NHI}$ and $C_{\rm high-\NHI}$ and the values of $\kappa_{\rm fit}$ and $dn_{\rm LLS,fit}/dX$ implied by this analytic approximation. Also listed are the 1-$\sigma$ perturbed values of $\beta_{\rm high-\NHI}$, the corresponding normalization $C_{\rm high-\NHI}$, and their implied $\kappa_{\rm fit}$ and $dn_{\rm LLS,fit}/dX$. Notably, the perturbed analytic parameters reproduce the opacity due to Ly$\alpha$ forest absorbers within the measurement errors. The incidence of LLSs is more sensitive to perturbed values of $\beta$ with offsets in the implied incidence of $\sim 1.5 \sigma$ from the observed incidence. As the incidence of LLSs is not used to constrain the power-law fit, the disagreement between these values is not surprising.

\begin{figure*}
\center
\includegraphics[width=0.5\textwidth]{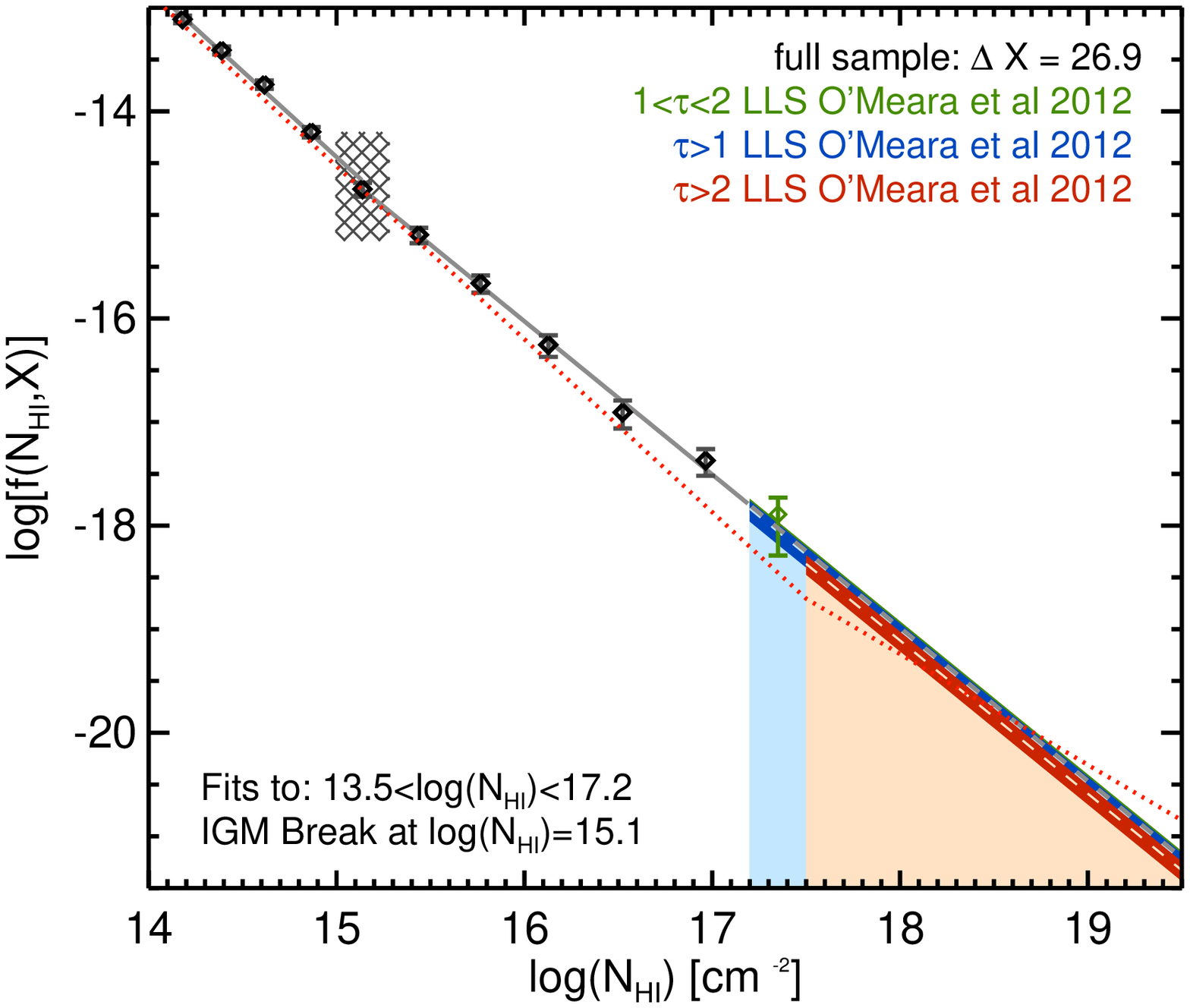}\includegraphics[width=0.5\textwidth]{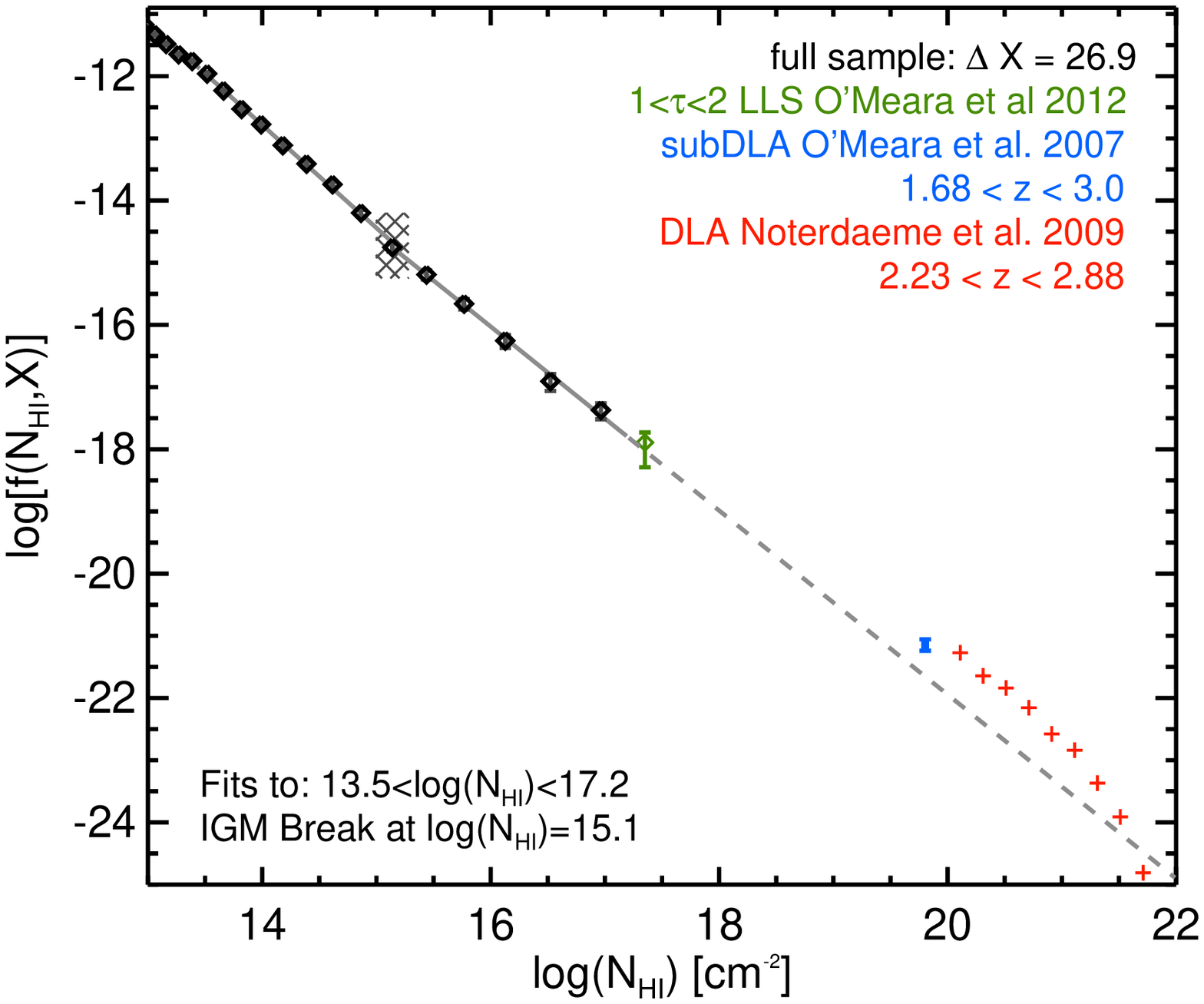}
\caption{\small Same at Figure \ref{broken_CGM} but only considering the IGM distribution. \textit{Left:}  The results of a recent survey for Lyman limit absorbers by \citet{ome12}. Note that these authors' constraints on $\tau_{912} > 1$ and $\tau_{912}>2$ LLS are well reproduced by the extrapolation of the power law slope selected for the high-\NHI\ IGM absorbers. Computing the difference between these values allows for an assessment of the number of LLSs with $1 < \tau_{912}<2$, here represented as the green data point.
The (red) dotted curve shows the fit to the frequency distribution proposed by \citet{ome12}. Note that this fit underpredicts the number of absorbers with $\log(\NHI)/\rm cm^{-2})<17.2$ as measured with the KBSS sample (black data points).
\textit{Right:} Including all currently known constraints at these redshifts for $f(N,X)$.  In green the $1 < \tau_{912}<2$ measurement from \citet{ome12}.  In blue are the subDLAs ($19.3< \log(\NHI/ \rm{cm}^{-2})<20.3$; $dn_{\rm subDLA}/dX = 0.110^{+0.026}_{-0.021}$)  from \citet{ome07} and in red are the DLAs ($\log(\NHI/ \rm{cm}^{-2})>20.3$; $dn_{\rm DLA}/dX \approx 0.05$) from \citet{not09}.
We note that the extrapolation of the intermediate-\NHI\ absorbers power law fit suggested by our Ly$\alpha$ forest data appears to provide a good fit to the data for $\log(\NHI/ \rm{cm}^{-2})\lesssim 18$. Above this \NHI, nature likely deviates significantly from this power law.}
\label{fNX_LLS_Omeara}
\end{figure*}

Figure \ref{fNX_LLS_Omeara} compares the IGM broken power law fit to the frequency of LLS with $\tau_{912} > 1$ and $\tau_{912} > 2$  from \citet{ome12} as described in \S \ref{LLS}. The blue triangle corresponds to $\tau_{912} > 1$  absorbers, and the red triangle represents the $\tau_{912} > 2$ absorbers. As in section \ref{LLS}, the normalization of the triangles are set by the measured value of $dn_{\rm LLS,data}/dX$ \textit{under the assumption that an extrapolation of the broken power law fit to the Ly$\alpha$ forest data holds within the LLS regime.} Again, the agreement between the extrapolated power-law fit (grey curve) and the blue and red triangles suggests that the analytic approximation to the frequency distribution reproduces the measure LLS statistics from  \citet{ome12}. The green point in Figure \ref{fNX_LLS_Omeara} shows the measured frequency of absorbers with $1<\tau_{912} < 2$  [$17.2< \log(\NHI/ \rm{cm}^{-2}) < 17.5$] calculated from the difference between the measured values of $dn_{\rm LLS,\tau>1}/dX$ and $dn_{\rm LLS,\tau>2}/dX$. 

 The power law fit to $f(N,X)$ suggested by \citet{ome12} is overplotted as the red dotted line for comparison. This curve would imply a value for the opacity from absorbers with $\log(\NHI/ \rm cm^{-2}) < 17.2$ of $\kappa_{\rm fit}=0.0022$, nearly a factor of two smaller than that measured from the KBSS sample ($\kappa_{\rm  data} = 0.0039\pm 0.0005$). 

There are likely some deviations in the detailed distribution of $f(\NHI,X)$ in the subDLA ($19.3< \log(\NHI/ \rm{cm}^{-2})<20.3$) and DLA ($\log(\NHI/ \rm{cm}^{-2})>20.3$) regimes as discussed in \citet{pro09} and \citet{ome12}. We show the current constraints on these regions of the frequency distribution in the right hand panel of Figure \ref{fNX_LLS_Omeara}. 
We chose to neglect the differences between the power law fit and observed constraints on $f(N,X)$. Holding the measured value of $dn_{\rm LLS}/dX$ fixed, such an assumption makes our measurements of the opacity from absorbers with $\NHI > 10^{17.2}$ \cm2\ a \textit{lower limit}. However, since the attenuation resulting from any system with $\log(\NHI/ \rm{cm}^{-2}) \gtrsim 18$ is similar, the detailed shape of the distribution for high-\NHI\ will introduce relatively small changes in the opacity so long as the total number of such systems is reproduced.

The opacity inferred from the power law fit to $f(N,X)$ is a lower limit for the following reason. If we hold $dn_{\rm LLS}/dX$ fixed at the observed value, this gives us a constraint on the number of systems with $\NHI > 10^{17.2}$ \cm2. The assumption of a given value of the power-law index, $\beta$, allows us to infer the opacity from a measurement of $dn_{\rm LLS}/dX$. Allowing $\beta$ to take a shallower value to match the plotted points of $f(N,X)$ for the subDLAs and DLAs only moves absorbers to higher \NHI\ and thus higher opacity compared to a model with a steeper power-law index. Therefore, our suggested opacities are likely a slight under-prediction of the true value. If we instead choose a flat value ($\beta=0$) we would find 9.7\% higher opacity from LLSs, suggesting the total uncertainty in the opacity from the assumed value of $\beta$ is significantly less than 10\%.

\begin{figure}
\center
\includegraphics[width=0.45\textwidth]{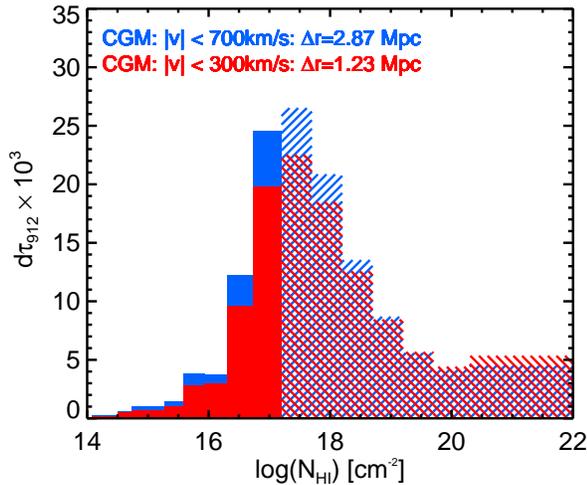}
\caption{\small The optical depth of the CGM to13.6 eV photons. The bins with $\log(\NHI/ \rm{cm}^{-2}) < 17.2$ (solid shading) are measured directly from the data. The bins corresponding to LLSs ($\log(\NHI/ \rm{cm}^{-2}) > 17.2$; hatched shadding) use the broken power law fit to the higher \NHI\ end of $f(\NHI,X)$ as shown in Figure \ref{broken_CGM}. The blue and red histograms show the total optical depth within one average CGM. This is equivalent to the opacity times the scale length $\Delta r_{\rm CGM}$. Note that the total optical depth in the 700 \kms\ CGM includes the opacity within the 300 \kms\ CGM but that there exists additional opacity due to absorbers with velocity offsets between $\pm300$ and $\pm700$ \kms.}
\label{tau_hist_LLS}
\end{figure}

\subsection{The Opacity of the IGM and CGM}
\label{opacity}

\begin{deluxetable*}{lcc|cccc|cccc}
\tablecaption{Opacity and Optical Depth of Absorbers of various \NHI}  
\tablewidth{0pt}
\tablehead{
\colhead{\NHI\ min} & \colhead{\NHI\ max} & \colhead{method} & \colhead{$\kappa_{\rm IGM}$} &  \colhead{$\tau_{\rm IGM}$} &  \colhead{$\kappa_{\rm CGM~700}$} &  \colhead{$\tau_{\rm  CGM~700} $} &  \colhead{$\kappa_{\rm IGM} $} &  \colhead{$\tau_{\rm IGM} $} &  \colhead{$\kappa_{\rm CGM~700}$} &  \colhead{$\tau_{\rm  CGM~700}$}\\
\colhead{} & \colhead{} & \colhead{} &  \colhead{$\times 10^3$ [pMpc$^{-1}$]} &  \colhead{$ \times 100 $} &  \colhead{$\times 10^3$ [pMpc$^{-1}$]} &  \colhead{$ \times 100 $} &  \colhead{$\times 10^3 $ [pMpc$^{-1}$]} &  \colhead{$ \times 100 $} &  \colhead{$\times 10^3 $ [pMpc$^{-1}$]} &  \colhead{$ \times 100 $}\\
\colhead{} & \colhead{} & \colhead{} & \multicolumn{4}{c}{Differential: \NHI\ min $<\log(\NHI/ \rm cm^{-2})<$\NHI\ max} &  \multicolumn{4}{c}{Cumulative: 12.55 $<\log(\NHI/ \rm cm^{-2})<$\NHI\ max}}
\startdata
12.55 & 12.60 & data &  0.001 &  0.02 &  0.001 &  0.00 &  0.001 &  0.02 &  0.001 &  0.00 \\ 
12.60 & 12.65 & data &  0.002 &  0.02 &  0.002 &  0.00 &  0.003 &  0.03 &  0.002 &  0.00 \\ 
12.65 & 12.71 & data &  0.002 &  0.02 &  0.002 &  0.00 &  0.005 &  0.06 &  0.004 &  0.00 \\ 
12.71 & 12.77 & data &  0.003 &  0.03 &  0.003 &  0.00 &  0.008 &  0.09 &  0.007 &  0.00 \\ 
12.77 & 12.85 & data &  0.004 &  0.04 &  0.003 &  0.00 &  0.012 &  0.13 &  0.009 &  0.00 \\ 
12.85 & 12.93 & data &  0.004 &  0.04 &  0.003 &  0.00 &  0.016 &  0.17 &  0.012 &  0.00 \\ 
12.93 & 13.01 & data &  0.006 &  0.07 &  0.006 &  0.00 &  0.022 &  0.24 &  0.018 &  0.01 \\ 
13.01 & 13.11 & data &  0.008 &  0.08 &  0.007 &  0.00 &  0.030 &  0.32 &  0.024 &  0.01 \\ 
13.11 & 13.21 & data &  0.009 &  0.10 &  0.010 &  0.00 &  0.039 &  0.42 &  0.034 &  0.01 \\ 
13.21 & 13.32 & data &  0.011 &  0.12 &  0.011 &  0.00 &  0.050 &  0.54 &  0.046 &  0.01 \\ 
13.32 & 13.45 & data &  0.017 &  0.18 &  0.016 &  0.00 &  0.067 &  0.72 &  0.061 &  0.02 \\ 
13.45 & 13.59 & data &  0.021 &  0.22 &  0.022 &  0.01 &  0.088 &  0.94 &  0.083 &  0.02 \\ 
13.59 & 13.74 & data &  0.024 &  0.25 &  0.032 &  0.01 &  0.112 &  1.20 &  0.116 &  0.03 \\ 
13.74 & 13.90 & data &  0.027 &  0.29 &  0.027 &  0.01 &  0.140 &  1.49 &  0.143 &  0.04 \\ 
13.90 & 14.08 & data &  0.038 &  0.40 &  0.053 &  0.02 &  0.178 &  1.89 &  0.195 &  0.06 \\ 
14.08 & 14.28 & data &  0.045 &  0.48 &  0.089 &  0.03 &  0.223 &  2.37 &  0.285 &  0.08 \\ 
14.28 & 14.50 & data &  0.066 &  0.70 &  0.073 &  0.02 &  0.289 &  3.07 &  0.358 &  0.10 \\ 
14.50 & 14.73 & data &  0.095 &  1.01 &  0.198 &  0.06 &  0.383 &  4.08 &  0.557 &  0.16 \\ 
14.73 & 15.00 & data &  0.115 &  1.23 &  0.353 &  0.10 &  0.499 &  5.31 &  0.910 &  0.26 \\ 
15.00 & 15.28 & data &  0.123 &  1.30 &  0.351 &  0.10 &  0.621 &  6.61 &  1.261 &  0.36 \\ 
15.28 & 15.60 & data &  0.202 &  2.14 &  0.511 &  0.15 &  0.823 &  8.75 &  1.772 &  0.51 \\ 
15.60 & 15.94 & data &  0.314 &  3.34 &  1.320 &  0.38 &  1.136 & 12.09 &  3.092 &  0.89 \\ 
15.94 & 16.32 & data &  0.471 &  5.01 &  1.316 &  0.38 &  1.607 & 17.10 &  4.408 &  1.26 \\ 
16.32 & 16.73 & data &  0.661 &  7.03 &  4.278 &  1.23 &  2.268 & 24.13 &  8.686 &  2.49 \\ 
16.73 & 17.20 & data &  1.695 & 18.03 &  8.571 &  2.46 &  3.963 & 42.17 & 17.257 &  4.95 \\ 
17.20 & 17.70 & fit  &  1.570 & 16.70 &  9.249 &  2.65 &  5.534 & 58.87 & 26.506 &  7.60 \\ 
17.70 & 18.20 & fit  &  1.111 & 11.82 &  7.273 &  2.09 &  6.645 & 70.69 & 33.779 &  9.68 \\ 
18.20 & 18.70 & fit  &  0.647 &  6.88 &  4.719 &  1.35 &  7.291 & 77.57 & 38.498 & 11.04 \\ 
18.70 & 19.20 & fit  &  0.372 &  3.96 &  3.030 &  0.87 &  7.664 & 81.53 & 41.527 & 11.91 \\ 
19.20 & 19.70 & fit  &  0.214 &  2.28 &  1.946 &  0.56 &  7.878 & 83.81 & 43.473 & 12.46 \\ 
19.70 & 20.30 & fit  &  0.141 &  1.50 &  1.439 &  0.41 &  8.019 & 85.31 & 44.912 & 12.88 \\ 
20.30 & 22.00 & fit  &  0.127 &  1.35 &  1.597 &  0.46 &  8.146 & 86.67 & 46.509 & 13.33 \\ 
 \enddata
 \tablenotetext{a}{Opacity calculated from the the data using equation \ref{kappa_data} for $12.55 \le \log(\NHI/ \rm{cm}^{-2}) \le 17.2$.}
     \label{opacity_tab}
\end{deluxetable*}

Under the assumptions outlined above, and with a parameterization of the frequency distribution that reproduces both the incidence of LLS and the opacity from intermediate-\NHI\ absorbers, we can now measure the full opacity of the IGM and CGM. We calculate the opacity from LLS absorbers with $\log(\NHI/ \rm{cm}^{-2}) > 17.2$ using equation \ref{kappa_fit}\footnote{Note that the detailed distribution of $\kappa$ per bin in \NHI\ would be affected by variation in $f(\NHI)$ at DLA and subDLA column densities. Because we adopt the extrapolation of the broken-power law fit to $f(N,X)$ to calculate the opacity, the values in individual bins of \NHI\ for $\log(\NHI/ \rm{cm}^{-2}) > 17.2$ are not well constrained.} and the opacity from intermediate-\NHI\  absorbers with  $\log(\NHI/ \rm{cm}^{-2}) < 17.2$ directly from the data using equation \ref{kappa_data} (as described in the beginning of Section \ref{LyC}). The differential opacity from bins in \NHI\ as well as the cumulative opacity (from absorbers with \NHI\ smaller than a given value), are given in Table \ref{opacity_tab}. The total opacity of the IGM is measured to be:
\begin{eqnarray}
\kappa_{\rm IGM} &=& \kappa_{\rm data}\left(\NHI<10^{17.2} \right) + \kappa_{\rm fit}\left( \NHI>10^{17.2}  \right) \\
\kappa_{\rm IGM} &=& 8.146 \times 10^{-3} ~\textrm{pMpc}^{-1}.
\end{eqnarray} 
Similarly,
\begin{eqnarray}
\kappa_{\rm CGM ~300~ km~ s^{-1}} &=& 94.4 \times 10^{-3} ~\textrm{pMpc}^{-1} \\
\kappa_{\rm CGM ~700~ km~ s^{-1}} &=&  46.5 \times 10^{-3} ~\textrm{pMpc}^{-1}.
\end{eqnarray}

Another relevant quantity is the integrated opacity over a specific scale length, i.e., the optical depth $\tau$. In particular, it is now possible to calculate the optical depth of the CGM of an individual galaxy. This quantity is particularly relevant for understanding the attenuation of ionizing flux emanating from a galaxies similar to those in the KBSS sample due to the CGM of the galaxy itself. Here we attempt to separate the probability that ionizing photons escape the ISM of a galaxy from the probability that photons, having already escaped the ISM, will also escape the enhanced absorption found within the CGM. 

Specifically, the KBSS data set constrains the opacity and optical depth of the CGM of galaxies at impact parameters greater than 50 kpc\footnote{This is because the closest galaxy to a QSO sightline in the KBSS sample is at a projected physical distance of 50 kpc.}. Henceforth, we consider the classical ``escape fraction'' to be the probability that an ionizing photon reaches a distance of 50 pkpc, while the CGM is defined to begin at a distance of 50 pkpc from the galaxy.  In order to calculate the optical depth, the scale length of interest must be assumed. 
For circumgalactic absorption, the velocity window associated with the CGM provides a natural scale length.
 As the relevant scale is along the line of sight, we use the distance, $\Delta r_{\rm CGM}$, associated with the line-of-sight velocity amplitude of the CGM:
\begin{equation}
\Delta r_{\rm CGM}=H^{-1}(z) d\rm{v}
\label{dr_cgm}
\end{equation}
where $H(z)$ is the Hubble constant evaluated at the mean redshift of the sample $\langle z \rangle = 2.4$ and $d\rm{v}$ is either 700 \kms\ or 300 \kms.  We note that the KBSS data set constrains the distribution of gas at transverse distances $>50$ pkpc from galaxies, not along the line of sight from galaxies. Here we assume that the gas distribution traced in the transverse plane well represents the gas distribution along the line of sight to a galaxy at distances $> 50$ kpc. Given the strong increase in \NHI\ at small galactocentric distances [as observed in \citet{gcr12}], the optical depth associated with the CGM at distances $> 50$ kpc is likely significantly smaller than the optical depth associated with the closer-in CGM. Therefore the measurements presented here might be thought of as a lower limit on CGM absorption.

In Figure \ref{tau_hist_LLS}, we compare the optical depth of the CGM ($\tau_{\rm CGM} = \kappa_{\rm CGM} \Delta r_{\rm CGM}$) for the 300 \kms, 300 pkpc CGM (red) and the 700 \kms, 300 pkpc CGM (blue). Figure \ref{tau_hist_LLS} also quantifies the optical depths from both forest absorbers ($\log(\NHI/ \rm{cm}^{-2}) < 17.2$) and LLS in the CGM. The forest absorber opacities and optical depths are measured directly from the data, while the LLS opacity and optical depths are inferred from an extrapolation of the high-\NHI\ fit to the data using the broken power law parameterization. The values for $\tau_{\rm CGM}$ and  $\kappa_{\rm CGM}$ in differential and cumulative \NHI\ bins are given in Table \ref{opacity_tab}. 

We find that the optical depth of the 700 \kms\ CGM is comparable to but slightly exceeds the optical depth in the 300 \kms\ CGM. This is as expected because the 300 \kms\ CGM includes the majority of the excess absorption; however, there is a small additional excess between 300 and 700 \kms, see \citet{gcr12}. For subsequent calculations, only the 700 \kms\ CGM will be considered since it captures the full optical depth of the CGM for $d > 50$ pkpc.

\begin{figure}
\center
\includegraphics[width=0.45\textwidth]{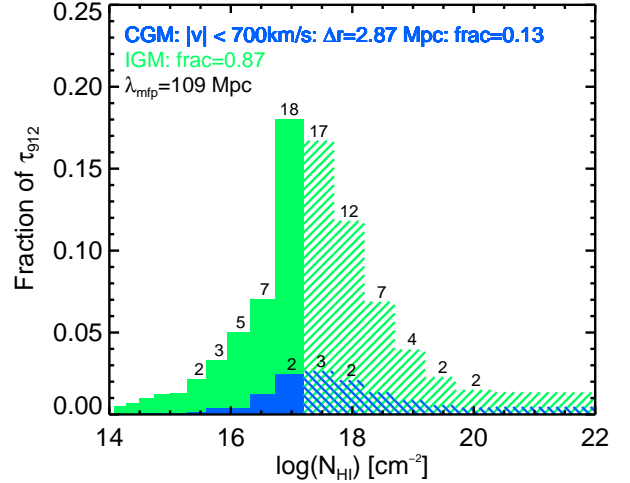}
\caption{\small The fraction of the optical depth within 1 mean free path contributed by absorbers within a given bin of \NHI\ in the IGM (green) or 700 \kms\ CGM (blue). The solid histograms represent measurements from the data converted into an optical depth using equations \ref{eqn_dt_dr} and \ref{kappa_data}. The hatched histograms represent values from the fit to the data extrapolated to higher values of \NHI\ and then converted into an optical depth using equations \ref{eqn_dt_dr} and \ref{kappa_fit}. The printed numbers give the percentage of the opacity due to that \NHI\ bin.}
\label{frac_tau}
\end{figure}

\begin{figure}
\center
\includegraphics[width=0.45\textwidth]{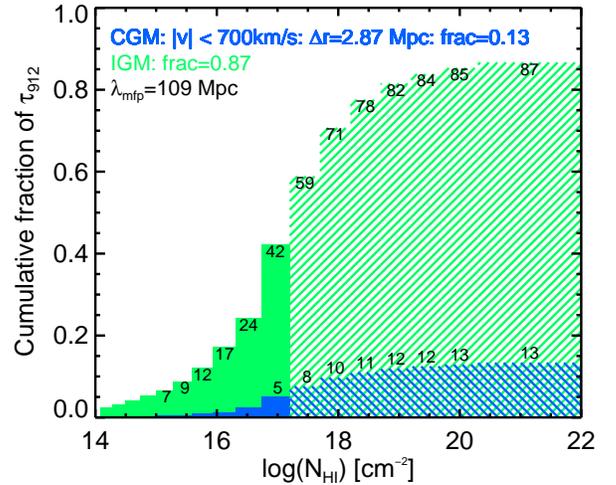}
\caption{\small The cumulative fraction of the optical depth within 1 mean free path contributed by absorbers below a given \NHI\ in the IGM (green) or 700 \kms\ CGM (blue). The solid histograms represent measurements from the data converted into an optical depth using equations \ref{eqn_dt_dr} and \ref{kappa_data}. The hatched histograms represent values from the fit to the data extrapolated to higher values of \NHI\ and then converted into an optical depth using equations \ref{eqn_dt_dr} and \ref{kappa_fit}. The printed numbers give the percentage of the opacity due to absorbers with \NHI\ less than or within the \NHI\ bin.}
\label{frac_tau_cum}
\end{figure}

\subsection{The Distribution of Optical Depths in the IGM and CGM}
\label{mfp_text}

With an estimate of the opacity from all \NHI\ absorbers, one can easily calculate the fractional contribution of the CGM vs the IGM, as well as the contribution from absorbers of different \NHI, to the mean free path. For this simplified calculation, we define the mean free path, \mfps, to be the average distance an ionizing photon with energy 1 Ry travels in a static universe before having a probability of $1-e^{-1}$ of being absorbed. More quantitatively: 
\begin{equation}
\mfps = \sum_i \Delta r_i
\end{equation}
where
\begin{equation}
\sum_i \kappa_i  \Delta r_i = 1
\end{equation}


Here we consider two possible mean free paths: (1) In the case of photons emanating from the ISM of a galaxy similar to those in our spectroscopic sample (see Section \ref{gal_fN}), they must first traverse the CGM of the galaxy in which they were formed and then the general IGM. Here we refer to this mean free path as \mfpsCGM .(2) The second case is the more general one and the mean free path that has been considered by previous authors - the mean free path of photons traveling in the IGM, a quantity relevant to photons once they have escaped the CGM of whichever structure in which they were formed. This distance is referred to as \mfpsIGM.

 In the calculation that follows, we consider the contributions to the opacity from the 700 \kms\ CGM as well as the IGM. Under these assumptions:
\begin{equation} 
\kappa_{\rm CGM} \Delta r_{\rm CGM} +  \kappa_{\rm IGM}(\mfpsCGM -  \Delta r_{\rm CGM})= 1.
\label{mfp_eqn}
\end{equation}
where $\Delta r_{\rm CGM} = 2.87$ pMpc as given in equation \ref{dr_cgm} using $dv = 700$ \kms, and $\kappa_{\rm IGM} = 0.0081$ pMpc$^{-1}$ and $\kappa_{\rm CGM}=0.047$ pMpc$^{-1}$ are the total opacities over all \NHI. Solving equation \ref{mfp_eqn}, we find 
\mfpsCGM $=109.3$ pMpc.
When we exclude the CGM component, we instead obtain 
\mfpsIGM$=122.8$ pMpc,
 a 12\% longer pathlength.

With \mfpsCGM\ known, we can then calculate the fractional contribution to the optical depth from each bin of \NHI\ for both the IGM and CGM as shown in Figure \ref{frac_tau} and Table \ref{opacity_tab} (middle columns). The cumulative version of this plot (i.e. the contribution to the optical depth for all \NHI $< N_i$) is given in Figure \ref{frac_tau_cum} and in Table \ref{opacity_tab}, right-hand columns. From these, we can see that the CGM accounts for 13\% of the opacity within a \mfpsCGM. Further, absorbers with $\log(\NHI/ \rm{cm}^{-2}) < 17.2$ contribute  48\% of the opacity to LyC photons.

In practice, the detailed distribution of opacity as a function of \NHI\ will be altered when photons of various energies are considered. The true \mfp\ of LyC photons will be non-negligibly affected by the redshifting of photons to lower energies, and by the presence of higher-energy radiation. For instance, 100 pMpc corresponds to  $\Delta z \approx 0.25$ at these redshifts. Photons emitted by a source at $\langle z \rangle =2.4$ with energies close to the Lyman Limit therefore will redshift to non-ionizing energies long before they are attenuated by $e$. 

In the next section, we consider a more rigorous calculation of \mfp\ including the aforementioned effects using Monte Carlo simulations of discrete lines of sight. As we will show, a more proper treatment of the physics does not substantially affect the value of \mfp.

\section{The Mean Free Path to Hydrogen-Ionizing Photons}

\label{mcText}

As we have shown in Section \ref{LyC}, the mean free path of LyC photons is a cosmological distance at $z=2.4$. As such, the consideration of photons with energies $>1$ Ry as well as inclusion of redshifting of photons out of ionizing frequencies are necessary in order to measure the true \mfp. It is also necessary to define very precisely what one means by the `mean free path,' including the frame of reference in which the distance is measured. Here we define the mean free path, \mfp, as the distance traveled from a source by a packet of photons before photons with energy of 1 Ry in the \textit{absorber's}\footnote{We consider the energy of the photon at the absorber's redshift because only photons with energy sufficient to ionize hydrogen when they intersect an absorber are relevant to \mfp. } frame are attenuated by an average factor of $e$. This distance is fundamentally related to a measurement of $\Delta z$, which we convert to the physical distance traveled by the photon through the expanding universe.

We perform a Monte Carlo (MC) simulation of the LyC absorption along many individual sightlines. For illustration, we begin with a discussion of MC runs with an emission redshift of $z=2.4$. 

\subsection{Monte Carlo Simulations}

\label{sims}

\begin{figure}
\center
\includegraphics[width=0.5\textwidth]{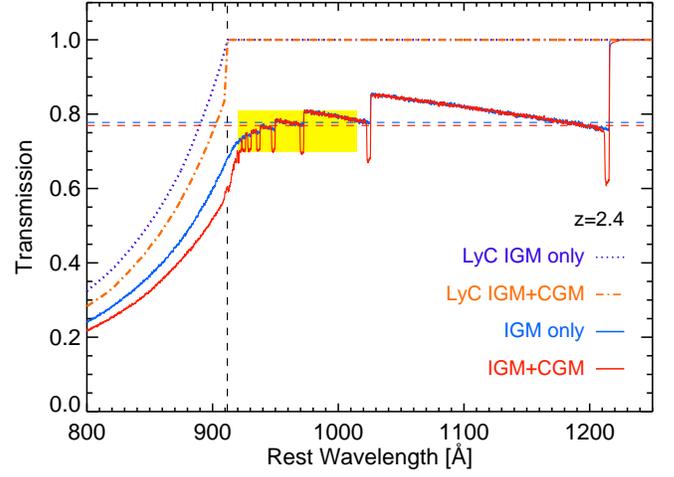}
\caption{\small The average transmission spectra from 10,000 runs of the Monte Carlo simulations of IGM only (purple and blue) and IGM+CGM(red and orange) absorption in the foreground of a $z=2.4$ emitter shown in the rest frame of the emitter. The solid (blue and red) curves show spectra using MC simulations that include line opacity. The dotted (purple) and dot-dashed (orange) curves show spectra from simulations only including continuum opacity. The highlighted yellow area marks the region between 920 -1015 \AA\ in the frame of the emitter which is used to renormalize the line opacity spectra in Figure \ref{NormMcSpec2.4}. The dashed horizontal lines mark the average value of the transmission in this region, $\langle 1-D_B \rangle$, for the spectra containing line opacity. The vertical dashed line marks the position of the Lyman limit in the frame of the emitter. The line absorption seen in the red spectrum is due to the Lyman series line opacity from high-\NHI\ CGM absorbers.}
\label{McSpec2.4}
\end{figure}

\begin{figure}
\center
\includegraphics[width=0.5\textwidth]{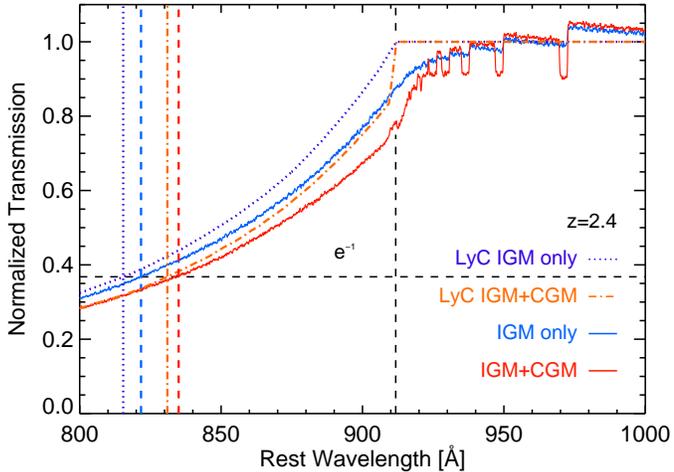}
\caption{\small The average normalized transmission below the Lyman limit of a $z=2.4$ emitter. The red and orange curves pertain to transmission including the CGM and the blue and purple curves pertain to IGM only transmissions. The dotted (purple) and dot-dashed (orange) lines show the same spectra as in Figure \ref{McSpec2.4} from the continuum opacity only runs of the MC code. The solid (red and blue) curves show the spectra including line opacity normalized by $\langle 1-D_B \rangle$. The black vertical dashed line marks the position of the Lyman limit in the frame of the emitter. The dashed horizontal line shows where the normalized transmission is equal to $e^{-1}$. The colored vertical lines mark the position where each of the spectra crosses the $e^{-1}$, signifying the position of one \mfp. These rest wavelengths are converted into physical distances corresponding to the \mfp\ as described in the text.}
\label{NormMcSpec2.4}
\end{figure}

To simulate absorption within intergalactic and circumgalactic space, we produced artificial absorption spectra whose absorption line distribution is matched to that observed in data at the same redshifts; such models are described in detail by \citet{sha06}. In this case, we produced 10,000 simulated lines of sight to sources with $z_{\rm em}=2.4$, according to a frequency distribution of absorbers set using best-fit parameters from the broken power law fit to $f(N,X)$ (Table \ref{broken_tab}). 

As was demonstrated in Section \ref{broken_text}, various fitting methods produce largely discrepant values for the parameters of the analytic form of $f(N,X)$. The actual opacity represented by the data can be measured to significantly higher precision then the parameters of the analytic form of $f(N,X)$. For this reason, using perturbed values of the power law parameters will result in significantly larger variation in the opacity along a line of sight than occur in the actual universe as measured by the data. As such, we use only the best-fit values of the power law fit to $f(N,X)$ in the Monte Carlo.


The simulated absorbers are distributed throughout the redshift range of interest in such a way that they reproduce the overall redshift distribution of measured absorbers. 
The redshift distribution of absorbers is typically parameterized as:
\begin{equation}
\frac{dn}{dz}=n_{0} (1+z)^\gamma.
\end{equation}
For the MC simulations, we have assumed that the Lyman-$\alpha$ forest absorbers with low-\NHI\ (those absorbers fit by the low-\NHI\ power law in the broken power law parameterization: $\log(\NHI/ \rm{cm}^{-2}) < 15.1$ for the IGM and $\log(\NHI/ \rm{cm}^{-2}) < 14.9$ for the CGM ) evolve with $\gamma=2.5$ \citep{kim02}. We have verified that the line sample presented in this paper exhibits evolution consistent with these measurements. All high-\NHI\ absorbers are assumed to have $\gamma=1.0$; the value for the LLS compilation reported in this paper (Section \ref{LLS}) measured through maximum likelihood estimation is $\gamma=0.93 \pm 0.29$. 

Briefly, we note that the redshift evolution for LLSs within the range $2\lesssim z \lesssim 3$ is well constrained by the data presented by \citet{rib11}, \citet{sar89}, and \citet{ste95} as discussed in Section \ref{LLS}. Measurements of the evolution in the line density at $z\gtrsim3$ \citep[see e.g.,][]{son10, pro10} suggest much steeper evolution with redshift (high values of $\gamma$). Because our Monte Carlo simulations only include absorption at $2\lesssim z \lesssim 3$, we use the value of $\gamma = 1$ as measured over the relevant redshift range and emphasize that the exact value of $\gamma$ does not significantly affect the quoted 
 \mfp\ within the redshift range $2< z< 3$. \footnote{Considered in the redshift range $2.0<z<2.8$, varying $\gamma$ by $1\sigma$ results in less than a 6\% change in the incidence of LLSs and therefore less than a 3\% change in \mfp.}
 However, we caution that the extrapolations of the trend of our \mfp\ measurements to $z>3$ is much more dependent on the chosen value of $\gamma$ and is therefore unlikely to predict the true value at higher redshift. 

An additional set of simulations was run to mimic lines of sight emanating from galaxies like those in the KBSS. In these simulations, \ion{H}{1} absorbers were added within 700 \kms\ of $z_{\rm em}$ according to the CGM broken power law fit to $f(N,X)$.  The remainder of the line of sight is drawn from the IGM distribution. These simulations produce the expected attenuation due to the CGM for sources within galaxies similar to those in our spectroscopic sample ($0.25< L/L^{*}<3$, $\langle z \rangle = 2.3$). 

In addition to the simulated forest spectra that include both line and continuum opacity, we created another set where the opacity from individual Lyman lines is not included, leaving only Lyman continuum absorption. To measure the \mfp, the principal concern is the fraction of photons absorbed while they are at ionizing energies. Once photons redshift out of the hydrogen-ionizing band, for the purpose of calculating the mean free path, their subsequent absorption or transmission is immaterial. For this reason, spectra with line and continuum opacity are useful for correcting observed objects for attenuation along the line of sight through the IGM (as will be discussed in Section \ref{line_continuum}), while the ``continuum absorption only'' spectra can be used to measure \mfp\ directly (as discussed in Section \ref{LyC_mfp}).

Figure \ref{McSpec2.4} shows the average of 10,000 realizations of each Monte Carlo run plotted in the rest-frame of the emitter. The solid curves show the runs with Lyman line opacity included, and the dotted and dash-dotted lines show the MC runs that only consider the continuum opacity. The various colors pertain to either IGM ONLY (blue and purple)  or IGM+CGM (red and orange) runs (note the presence of Ly$\alpha$, $\beta$, $\gamma$, etc. absorption due to the CGM of the emitter in the red spectrum). Note that at the Lyman limit, the spectra that include CGM absorption show a sharper drop in transmission. This is due to LyC absorption from high-\NHI\ absorbers found in the circumgalactic gas.

\subsection{Measuring \mfp}

\label{LyC_mfp}

To measure \mfp\ from the Monte Carlo calculation, we use the spectra that include continuum absorption only.\footnote{Line opacity only reduces the transmission of non-ionizing photons which are unimportant for consideration of \mfp.} We search the average spectrum for the location where the LyC only spectra reach a transmission equivalent to $e^{-1}$, as marked by the vertical dash-dotted lines in Figure \ref{NormMcSpec2.4}. The rest wavelengths from Figure \ref{NormMcSpec2.4} can be converted into the physical distance traveled by a photon between $z_{\rm em}=2.4$, the emission redshift of the spectra, and $z_{\rm mfp}$ where:
\begin{equation}
z_{\rm mfp}= \frac{\lambda_{e} - \lambda_{\rm LL}}{\lambda_{\rm LL}}(1+z_{\rm em})+z_{\rm em}
\label{phys_dist}
\end{equation}
where $\lambda_{\rm LL} = 911.75$ \AA\ is the wavelength of the Lyman limit and $\lambda_e$ is the rest wavelength of the spectrum at which the transmission equals $e^{-1}$. To calculate \mfp, we then compute the physical distance traveled by the photon through the expanding universe, given by the integral over the proper line element:
\begin{equation}
\mfp \equiv \int^{z_{\rm em}}_{z_{\rm mfp}} \frac{1}{1+z}\frac{c}{H(z)} dz
\label{dphys}
\end{equation}
where $c$ is the speed of light and 
\begin{equation}
H(z)=H_0\sqrt{\Omega_\Lambda + \Omega_{\rm m}(1+z)^3}.
\end{equation}

Using the averaged LyC only spectra shown in Figure \ref{NormMcSpec2.4}, we calculate the \mfp\ through the IGM assuming $z_{\rm em}=2.4$ to be 147 pMpc (or $\Delta z=0.359$). If instead the CGM is included (as would be the case for LyC photons emanating from a galaxy), the estimated \mfp\ falls to 120 pMpc ($\Delta z=0.301$),  $\sim$20\% shorter. Note that the values of \mfp\ found in this sections using the MC simulations are similar to those estimated using the analytic approximations in Section \ref{LyC} (\mfpsIGM$=122.8$ pMpc, \mfps $=109.3$ pMpc). This provides confidence that the conclusions of Section \ref{LyC} hold even when a more rigorous treatment is considered.

\begin{figure*}
\center
\includegraphics[width=0.9\textwidth]{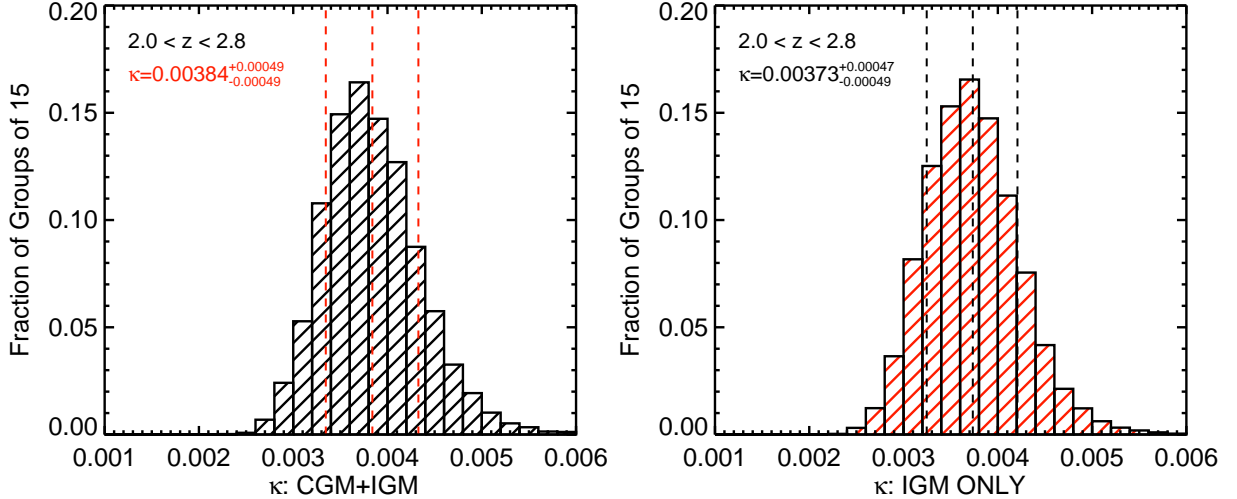}
\caption{\small The variation in the measured opacity in groups of 15 sightlines due to all absorbers with $\log(\NHI/ \rm{cm}^{-2}) < 17.2$. The simulations used in this plot are those with emission redshift $z=2.8$. These show the results of 10,000 jack-knife tests in which 15 spectra were randomly drawn from the ``LyC ONLY'' Monte Carlo runs in which only absorbers with $\log(\NHI/ \rm cm^{-2})<17.2$ were included. See Section \ref{mfp_error}. The dotted vertical lines mark the $1-\sigma$ uncertainties in $\kappa$. }
\label{subLLS_var}
\end{figure*}

\subsection{Computing the mean free path uncertainties}

\label{mfp_error}

To calculate the uncertainties in the \mfp\ measurement made in Sections \ref{LyC_mfp}, we must estimate the error in the opacity that originates from the Ly$\alpha$ forest absorbers ($\log(\NHI/ \rm{cm}^{-2}) < 17.2$) and from LLSs ($\log(\NHI/ \rm{cm}^{-2}) > 17.2$). The errors associated with each must be estimated separately because the constraints on the incidence of these absorbers come from different surveys. 

To estimate the uncertainty associated with LLSs, we return to the measured value of $dn_{\rm LLS}/dX = 0.52 \pm 0.08$ (Section \ref{LyC}). Using the method outlined in Section \ref{LyC}, we employ a fit to the frequency distribution to calculate the opacity, $\kappa$. Once again, we assume that an extrapolation of the broken power law fit to the slope of the high-\NHI\ absorbers holds in the LLS regime. We then calculate the normalization of the frequency distribution that reproduces $dn_{\rm LLS}/dX$ and its upper and lower limits. These normalizations and the assumed slope can then be used to measure $\kappa_{\rm LLS}$ and an approximate error. We find $\kappa_{\rm LLS} = 0.0043 \pm 0.0006$ Mpc$^{-1}$, a 15\% uncertainty. If we allow the value of $\beta$ to vary within the $1-\sigma$ confidence region (see Table \ref{beta_marg}), the uncertainty in our estimate of $\kappa_{\rm LLS}$ would be larger, $\sim 20-25\%$.

To estimate the \mfp\ uncertainties from absorbers with $\log(\NHI/ \rm{cm}^{-2})<17.2$ we employ the measurement of $\kappa_{\rm data}$ from Section \ref{LyC}. We thus assume that the uncertainty is dominated by sample variance rather than the uncertainties in the model. However, before using this value, we verify that the opacity (and its uncertainty) from $\log(\NHI/ \rm{cm}^{-2}) < 17.2$ absorbers measured from the data are reproduced in the Monte Carlo simulations using the assumed power-law fits. 

To estimate the opacity in the MC runs from forest absorbers, we ran another set of MC simulations of LyC only opacity. In this case, we included only absorbers with  $\log(\NHI/ \rm{cm}^{-2}) < 17.2$, excluding the contribution from LLSs. Using the ensemble of model spectra, we performed a jack-knife test with a sample size of 15 sightlines, designed to match the number of lines of sight in the KBSS QSO sample. Each set of 15 randomly-drawn spectra was averaged and the maximum attenuation with respect to the continuum was measured and recorded. The rest-frame wavelength at the position of maximum attenuation was converted into a proper distance at the emission redshift using equation \ref{dphys}. The average value of $\kappa$ along those 15 lines of sight was taken to be the natural logarithm of the minimum flux divided by physical distance. The redshift range used in the simulations is $2.0 < z < 2.8$ in order to reproduce the range of the KBSS QSO sightline sample. 

The jack-knife technique was applied 10,000 times to produce a distribution of the forest opacity, $\kappa_{\rm for}$, as shown in Figure \ref{subLLS_var}. An estimate for the error in $\kappa_{\rm for}$ was then derived from the moments of the distribution. The resulting value of $\kappa_{\rm for}=0.0037 \pm 0.0005$  is in good agreement with $\kappa_{\rm data}=0.0039\pm 0.0005$ calculated in Section \ref{LyC}, providing confidence that the MC simulations well reproduce both the mean value and the dispersion in $\kappa_{\rm for}$ for a sample of the same size as the KBSS.


With an estimate of $\kappa_{\rm LLS}$, $\sigma_{\kappa, \rm LLS}$ as well as $\kappa_{\rm for}$ and $\sigma_{\kappa, \rm for}$, it is straightforward to calculate the error in the \mfp. If
\begin{equation}
\mfp = \frac{1}{\kappa_{\rm LLS} + \kappa_{\rm for} }
\end{equation}
(see equation \ref{mfp_eqn}) then the uncertainty on \mfp\ is
\begin{equation}
\sigma_{\rm mfp} = \frac{\sqrt{\sigma_{\kappa, \rm LLS}^2 + \sigma_{\kappa, \rm for}^2  }}{(\kappa_{\rm LLS} + \kappa_{\rm for} )^2}
\end{equation}
where $\sigma_{\kappa, \rm LLS}$ and $\sigma_{\kappa, \rm for}$ are the uncertainties of $\kappa_{\rm LLS}$ and $\kappa_{\rm for}$ respectively.
Employing these equations, we find a typical uncertainty in \mfp\ of $\sim$ 15 Mpc, which we adopt for both the IGM and IMG+CGM \mfp\ measurements. Since the contribution of LLSs to the opacity in the CGM is poorly constrained, the error on the CGM measurements is likely somewhat larger; however since the CGM contributes $< 20\%$ of the opacity, the effect on the net uncertainty is modest. We adopt this uncertainty for the measurement of \mfp\ at the mean redshift of the sample.

\subsection{Simulated sightlines with line and continuum opacity}

\label{line_continuum}

\begin{figure*}
\centerline{
\includegraphics[width=0.45\textwidth]{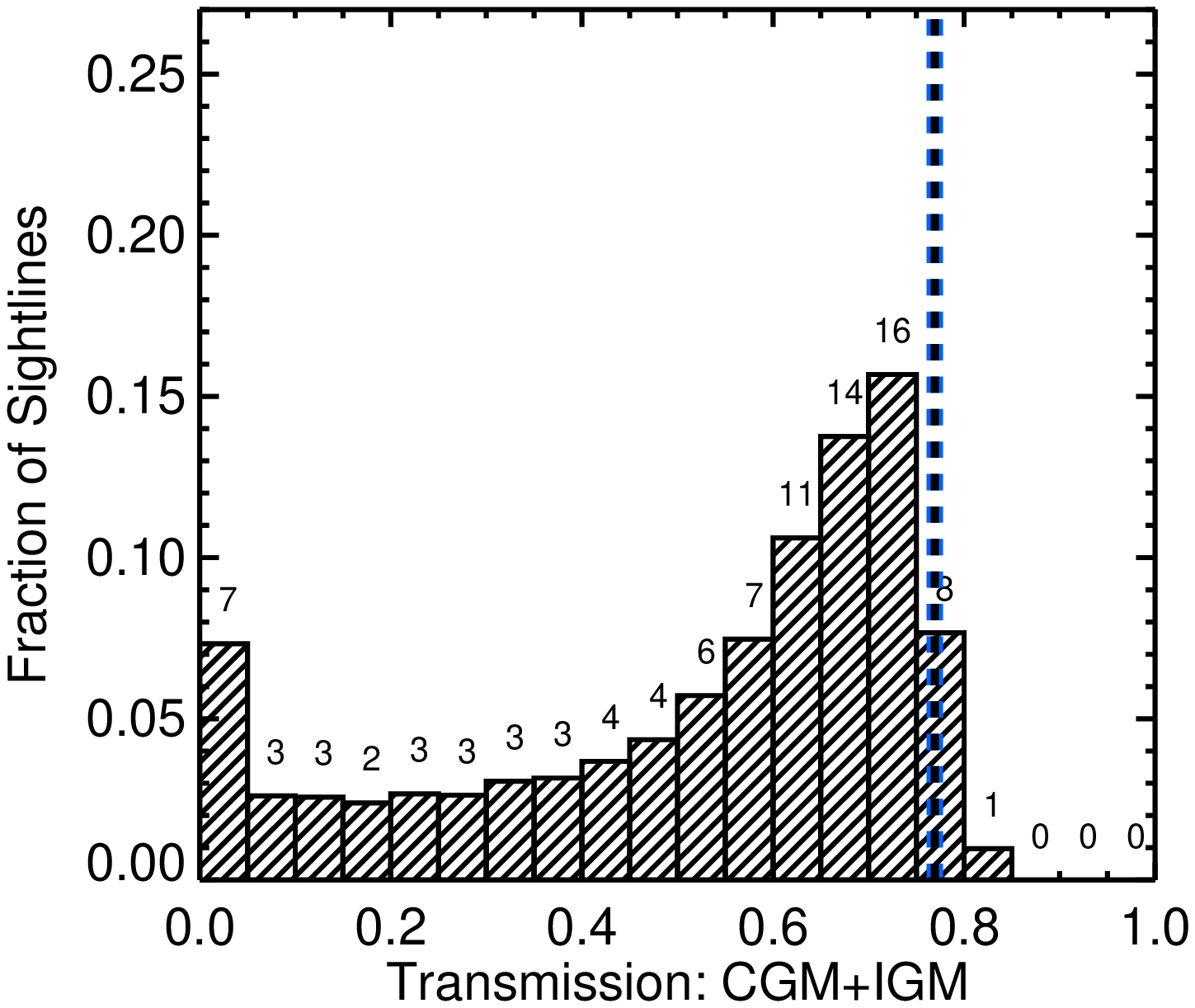}\includegraphics[width=0.45\textwidth]{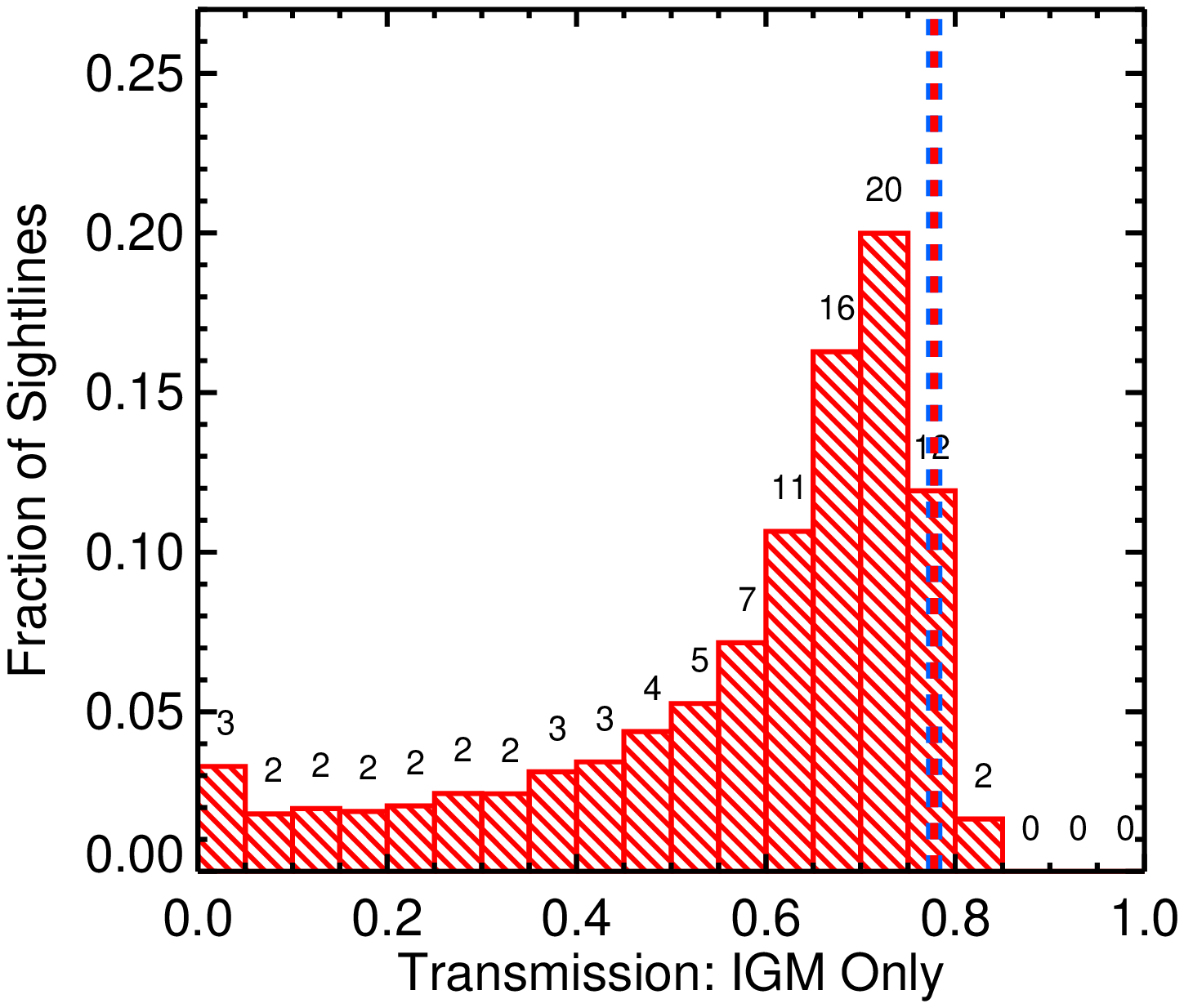}}
\caption{\small The distribution of transmissions from each sightline averaged over the band pass from 880-910 \AA\ for the the IGM+CGM Monte Carlo simulations (left panel) and IGM only Monte Carlo simulations (right panel) that include line opacity. The simulations used for these figures are those with an emission redshift of $z=2.4$. The (blue) dashed vertical line marks the average transmission in the Ly$\beta, \gamma, +$ forest $\langle 1-D_B \rangle$. }
\label{MCtrans}
\end{figure*}

For some applications, it is more useful to know the expected attenuation along an ensemble of sight lines including line opacity. For instance, in studies of the escape fraction of LyC photons (\citealt{sha06,iwa09,bri10,sia10,nes11}; Steidel et al. in prep) a correction must be made for the fraction of the attenuation at a given observed wavelength due to (non-ionizing) line blanketing. For these applications, our models including CGM attenuation are ideal. The sample of galaxies used in our CGM study have impact parameters of $50-300$ pkpc from the QSO line of sight, and so our MC simulations which include opacity from the CGM effectively model the typical gas distribution at radii larger than 50 kpc - and hence the attenuation suffered by photons after they leave the \textit{ISM} of the galaxy, but before they leave the large gas densities associated with the CGM.

Using the sets of simulated spectra with $z_{\rm em}=2.4$ that include Lyman line opacity for the IGM ONLY and the IGM+CGM, we measure the distribution of transmission of emitted 880-910~\AA\ photons through the IGM and the IGM + CGM; see Figure \ref{MCtrans}. The transmission is generally lower along the IGM+CGM sight lines. The peak in the transmission PDF near $\sim0.75$ corresponds to the average transmission within the Ly$\beta$ forest which is $<1$ due to line blanketing from both higher Lyman series lines as well as Ly$\alpha$ absorption from lower redshift.

\begin{deluxetable}{lcccc}
\tablecaption{Comparison of the \mfp\ derived from the LyC and the renormalized forest spectra}  
\tablewidth{0pt}
\tablehead{
\colhead{$z_{\rm em}$} & \colhead{\mfp\ LyC} & \colhead{\mfp $\langle1-D_B\rangle$} & \colhead{\mfp\ LyC} & \colhead{\mfp $\langle1-D_B\rangle$}\\
\colhead{} & \colhead{[pMpc]}  & \colhead{[pMpc]}  & \colhead{[pMpc]}  & \colhead{[pMpc]} \\
\colhead{} & \colhead{IGM ONLY} & \colhead{IGM ONLY} & \colhead{IGM+CGM} & \colhead{IGM+CGM}
}
\startdata
2.2 & 191.8 & 185.9 & 160.2 & 151.4\\
2.4 & 146.5 & 135.7 & 120.0 & 113.4\\
2.6 & 115.7 & 106.8 &  93.7 &  ~85.6\\
2.8 &  91.9 &  85.1 &  73.5 &  ~66.0
 \enddata
     \label{mfp_tab2}
\end{deluxetable}

In Figure \ref{NormMcSpec2.4}, the solid curves show the average spectrum of 10,000 MC runs that include line opacity. We use $\langle1-D_B\rangle$, the average transmission in the region between 920 -1015 \AA\ in the rest frame of the source (the region shown in the yellow box in Figure \ref{McSpec2.4}; \citealt{oke82})  to ``re-normalize'' the continuum against which the LyC opacity will be measured. Using the renormalized spectrum as an approximation to the ``LyC only'' MC spectra and measuring \mfp\ as described in Section \ref{LyC_mfp},  leads to \mfp\ values $5-10$\% smaller than the ``true'' values. Table \ref{mfp_tab2} compares the results of the MC simulations including line opacity (columns labeled $\langle1-D_B\rangle$) with those based on the continuum only spectra (columns labeled LyC).

\subsection{Comparison with previous \mfp\ measurements}

In the previous sections, we derived the value of the mean free path \mfp\ to LyC photons traveling through the IGM to be: 
\begin{equation}
\textrm{IGM ONLY: } \mfp (z_{\rm em}=2.4) = 147 \pm 15 ~\textrm{Mpc}.
\end{equation}
If instead, we assume such LyC photons emanate from galaxies similar to one of those in our spectroscopic sample, we find a value for \mfp\ including the CGM opacity to be:
\begin{equation}
\textrm{IGM+CGM: } \mfp (z_{\rm em}=2.4) = 121 \pm 15 ~\textrm{Mpc}.
\end{equation}
  
 One important advance of the estimate of \mfp\ made in this paper is that all parts of the column density distribution have now been measured at the same mean redshift. This has not been possible previously due to the difficulty of collecting appropriate statistical samples for the various column density regions of the frequency distribution. 
 
\begin{figure}
\center
\includegraphics[width=0.5\textwidth]{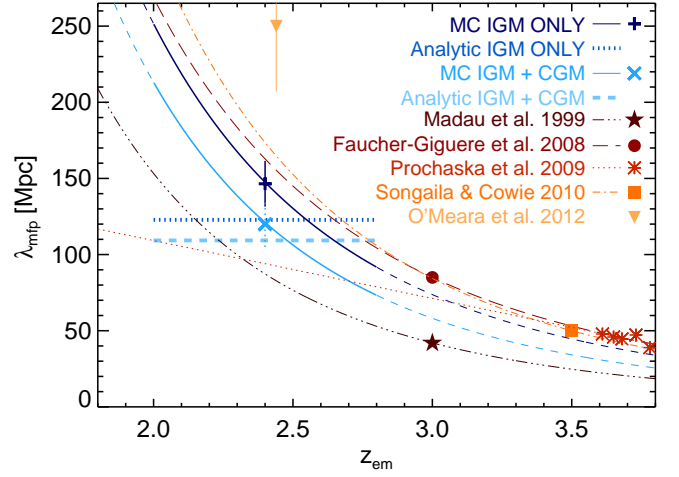}
\caption{\small The \mfp\ as a function of $z_{\rm em}$ including previous estimates of the \mfp\ from the literature. The blue points and curves refer to values measured in this paper whereas estimates from the literature are plotted in shades of red. The blue cross and X and their corresponding curves show the values of \mfp\ calculated from the MC spectra for the IGM ONLY and IGM+CGM runs respectively.  The dotted and dashed blue horizontal line show the results of the analytic calculation in Section \ref{LyC} excluding and including the CGM respectively.
 The dark red filled star and the dash-dotted curve that intersects it are the suggested values of \mfp\ from \citet{mad99}. The red filled circle and dashed curve are the results of \citet{fau08}. \citet{pro09} suggested the values shown with the red/orange asterisks and the dotted line. The measurements of \citet{son10}  are shown in the orange filled square and the dash-dotted curve that intersects it. The results of \citet{ome12} are shown in the orange filled triangle.}
\label{mfp_z_compare}
\end{figure}

 In this section, we compare these values to previous estimates from the literature. One caveat is that the measurements made in this paper typically differ in mean redshift from the samples on which previous \mfp\ measurements have relied. As such, an extrapolation of the literature measurements or our own measurements are generally necessary in order to compare the results, and therefore these comparisons are more dependent on the value of $\gamma$ assumed.  Nevertheless, in this section we compare to previous measurements for completeness. Figure \ref{mfp_z_compare} gives a graphical summary of the results of this comparison. 
 
  In general, the differences between our IGM ONLY \mfp\ results and those of other authors (especially those measured at similar $z$) underscore the degree to which \mfp\ is sensitive to changes in $f(N,X)$, highlighting the importance of the precise measurement of the intermediate \NHI\ systems made here. 
  
   One other caveat is the small degree of incompleteness which may be present in the $z\lesssim2.4$, $\log(\NHI/\rm cm^{-2})\gtrsim15.5$ portion of our absorber sample. As described in Appendix \ref{App_fN}, this bias will only act to \textit{shorten} the inferred value of \mfp, pushing it to values farther below those inferred by other authors.

 The mean free path and its evolution with redshift are generally parameterized as:
\begin{equation}
\mfp=\lambda_{{\rm mfp},0} \left(1+z\right)^{\xi}
\end{equation}
where $\xi$ captures the cosmology (as parameterized in equation \ref{dphys}) as well as the evolution in the number of absorbers as a function of redshift, $dn/dz$ (described in Section \ref{sims}). For the Monte Carlo simulations described in this section, the redshift evolution of \mfp\ is fixed by the assumed value of $\gamma$. We emphasize that the redshift evolution is only constrained by data with $2.0 \lesssim z \lesssim 3.0$, and as such the extrapolation of our results above to $z>>3$ is ill-advised. 

The redshift evolution of \mfp, parameterized as a function of the emission redshift of the QSO, $z_{\rm em}$, returned by the MC simulations is represented by the solid blue curves in Figure \ref{mfp_z_compare} and is given by:
\begin{equation}
\textrm{IGM ONLY: } \mfp (z_{\rm em}) = 147 \left( \frac{1+z_{\rm em}}{3.4}\right)^{-4.3} \textrm{pMpc}
\end{equation}
and
\begin{equation}
\textrm{IGM+CGM: } \mfp (z_{\rm em}) = 121 \left( \frac{1+z_{\rm em}}{3.4}\right)^{-4.5} \textrm{pMpc}.
\end{equation}



The first estimate of \mfp\ was made by \citet{mad99} who found:
\begin{equation}
\mfp = 42 \left( \frac{1+z}{4.0}  \right)^{-4.5} \textrm{Mpc} = 87 \left( \frac{1+z}{3.4}  \right)^{-4.5} \textrm{Mpc}
\end{equation}
using a simple approximation for $f(N,X)$ with $\beta=1.5$ and $\gamma=2$ where their Einstein-de Sitter cosmology has been converted to the cosmology used in this paper. 

\citet{fau08}  published an updated estimate to the \mfp\ assuming $\beta=1.39$ as measured by \citet{mis07}. 
\citet{mis07} measured the frequency distribution close to high-\NHI\ absorbers, and thus their sample is similar in slope to our CGM $f(N,X)$ distribution. However, the normalization chosen by \citet{fau08} was computed using LLS statistics for the full IGM, thereby closely matching our LLS points but somewhat under-predicting the incidence of lower-\NHI\ absorbers resulting in slightly larger values of \mfp\ at a given $z$: 
\begin{equation}
\mfp = 85 \left( \frac{1+z}{4}  \right)^{-4} \textrm{Mpc} = 160 \left( \frac{1+z}{3.4}  \right)^{-4} \textrm{Mpc}
\end{equation}

\citet{pro09} estimated \mfp\ using stacked QSO spectra to measure the LyC opacity. Notably, such a method would include any contribution from the CGM of their QSOs. They found 
\begin{equation}
\mfp=48.4 -38.0(z-3.6)~ \textrm{Mpc}  = 94-38.0(z-2.4)~ \textrm{Mpc}
\end{equation}
somewhat lower than our measurement at $z=2.4$ and with very different evolution with redshift than indicated by our data. 

\citet{son10} revisited the mean free path calculation after compiling a large sample of $z<1$ and $z>4$ LLS. They found
\begin{equation}
\mfp = 50 \left( \frac{1+z}{4.5}  \right)^{-4.44^{+0.36}_{-0.32}} \textrm{Mpc} = 170 \left( \frac{1+z}{3.4}  \right)^{-4.44^{+0.36}_{-0.32}} \textrm{Mpc} 
\end{equation}
using  $\beta=1.3$. 
They also offer a variety of normalizations for different $\beta$ values, including 130 Mpc for $\beta=1.5$ at $z=2.4$, in general agreement with our measured ``IGM ONLY'' value of \mfp.

Recently  \citet{ome12} presented a calculation of \mfp\ using their new measurements of the incidence of LLSs at $z\sim2$. Their estimates of the incidence of absorbers with $\log(\NHI/ \rm{cm}^{-2}) < 17.2$ leads to a lower value for the opacity as described in Section \ref{broken_text}. They found
\begin{equation}
\mfp = 250 \pm 43~ \textrm{Mpc}
\end{equation}
at $z=2.44$, nearly a factor of 2 larger than our estimate and significantly larger than the extrapolations of others' measurements at higher redshifts.

We note that for a steeper evolution in redshift as found in \citet{son10} and \citet{pro10}, the general agreement between the extrapolation of the presented measurements at $\langle z \rangle = 2.4$ to $z>3$ would result in a more discrepant value significantly below the measurements at $z>3.5$ by \citet{son10} and \citet{pro09}.

To summarize, the calculation described in Section \ref{mcText} offers three significant improvements over previous measurements. First, all portions of $f(N,X)$ have been measured at the same mean redshift, avoiding the need to extrapolate any part of $f(N,X)$ to a different redshift using uncertain values of $\gamma$. A second advance is the accuracy of our Voigt profile fitting technique, achieved by including higher-order Lyman series transitions, yielding the best measurement of the power law slope, $\beta$, for absorbers with saturated Ly$\alpha$ profiles,  [$14< \log(\NHI/ \rm{cm}^{-2}) < 17.2$]. Because such absorbers account for $\sim$half of the opacity to LyC photons in the IGM, underestimating their contribution would result in significantly underestimating the total opacity of the IGM. 

The third major improvement with respect to previous studies is the inclusion of absorption from the CGM which again decreases the measured value of \mfp. If LyC photons are presumed to originate within galaxies similar to those in the KBSS sample, then every hydrogen-ionizing photon must first propagate through the CGM. As shown by \citet{gcr12}, regions of the IGM near galaxies have a much higher incidence of high-\NHI\ absorbers than the average IGM with the effect (again) of reducing \mfp\ relative to previous measurements. 

\subsection{Implications of the higher opacity of the IGM+CGM}

A shorter mean free path has implications for the emissivity and demographics of ionizing sources at high redshift, as well as for calculations of the metagalactic UV radiation field. 
Briefly, if one considers the specific intensity of the UV background ($J_\nu$, see e.g. \citealt{sco00}) or the photoionization rate ($\Gamma$, see e.g. \citealt{fau08})  to be well determined, a shorter mean free path directly implies a requirement for correspondingly larger ionizing emissivity of sources ($\epsilon_\nu$). Following \citet{fau08}, 
\begin{equation}
J_\nu(z) \approx \frac{1}{4\pi} \mfp(\nu,z)~ \epsilon_\nu(z)
\end{equation}
and 
\begin{equation}
\Gamma \propto J_\nu.
\end{equation}
Conversely, for a measured value of $\epsilon_\nu$ from a given population of sources, a shorter \mfp\ implies a smaller contribution to both the specific intensity and the photoionization rate ($\Gamma$). Higher intergalactic opacity also requires that the background be produced by sources within a smaller volume, possibly leading to greater spatial variation in the radiation field intensity.

LyC escape fraction studies have often found very few detections of ionizing photons escaping from galaxies (\citealt{sha06,bri10,sia10}; however see  \citealt{iwa09}, \citealt{nes11}, and Steidel et al. in prep). For LyC studies of galaxies at high-$z$, the inclusion of the CGM is an important effect. Considering Figure \ref{MCtrans}, note that the number of galaxies with $\sim$zero transmission increases by a factor of 2 when the CGM is included. If the redshift evolution of the CGM is similar to that measured for the IGM, the effect of the CGM will become more pronounced at higher-$z$ where the incidence of high-\NHI\ systems in the CGM will be much higher. 

Higher IGM and CGM opacities mean that there may be more sources contributing to the metagalactic UV radiation field than are implied by the number of actual LyC detections; on the other hand, the assumption of a
$\lambda_{\rm mfp}$ that is too large may cause one to over-estimate the contribution to $J_{\nu}$ or $\Gamma$
made by a population of detected LyC sources. The latter effect may reduce the tension between the apparent over-production of ionizing photons by observed LBGs at $z \sim 3$ \citep{nes11} relative to estimates of the total photoionization rate produced by all sources \citep{bol05,fau08}.

\section{Summary}

\label{con}

Using a sample of 15 high-S/N, high-resolution QSO spectra drawn from the KBSS, we have produced the largest catalog of \ion{H}{1} absorbers fit with Voigt profiles to Ly$\alpha$ and at least one higher order Lyman series transition. This analysis enables the first statistically rigorous measurements of the frequency distribution of \ion{H}{1} absorbers with $14 \lesssim \log(\NHI/ \rm{cm}^{-2}) \lesssim 17$. In Section \ref{fN_text}, we showed that the frequency of absorbers as a function of \NHI\ is well-parameterized by a single power-law from $13.5 < \log(\NHI/ \rm{cm}^{-2}) < 17.2$ with a maximum likelihood index of $\beta = 1.65 \pm 0.02$ and a normalization per unit pathlength of $\log(C_{\rm HI}) = 10.32$. 

In Section \ref{gal_fN}, we measured the frequency distribution within 300 pkpc and both 300 \kms\ and 700 \kms\ of galaxies in the KBSS sample. We showed that the frequency of absorbers near galaxies is significantly higher and that the power-law index is shallower compared with that of the IGM, meaning there are disproportionately more high-\NHI\ systems than low. These findings are discussed further by \citet{gcr12}.

Section \ref{LyC} examined statistically the total opacity of the IGM and CGM to hydrogen-ionizing Lyman continuum (LyC) photons. This section presented measurements of the opacity due to absorbers with $\log(\NHI/ \rm{cm}^{-2}) < 17.2$ with direct constraints from the data. We also found that a single MLE power-law fit over the full range of \NHI\ does not well reproduce this opacity, and also fails to reproduce the incidence of LLSs. We therefore adopted a broken power law fit that reproduces both the opacity measured within the data from absorbers with $\log(\NHI/ \rm{cm}^{-2}) < 17.2$ as well as the observed incidence of LLSs. The parameters of these fits can be found in Table \ref{broken_tab}.

We also measured the fractional LyC opacity in bins of \NHI. We found that 48\% of the opacity within one mean free path, \mfp, is contributed by absorbers with \NHI $<10^{17.2}$ \cm2, the vast majority of which results from absorbers with $14 < \log(\NHI/ \rm cm^{-2}) < 17.2$ whose frequency was poorly measured prior to this work. 

In Section \ref{mcText}, we used Monte Carlo simulations to measure the value of the mean free path (\mfp) of LyC photons at $\langle z \rangle = 2.4$. We measured both the \mfp\ of photons through the IGM (as has been considered many previous times) as well as the \mfp\ of photons formed in galaxies similar to those in the KBSS -  photons that must first transverse the CGM of these galaxies before reaching the lower-opacity IGM.  We found 
$\mfp(z=2.4) = 147 \pm 15$ Mpc for simulations including only IGM opacity and 
$\mfp (z=2.4) = 121 \pm 15$  Mpc when we include the CGM.  These values of \mfp\ are lower than most previous estimates, which have not included the CGM and have generally underestimated the contribution of intermediate-\NHI\ absorbers. 
We also quantified the effect of the CGM on the distribution of transmission through random sight lines, relevant to measurements of the escape fractions of LyC photons from such galactic sources at high-$z$. 

We note that the \mfp\ value including attenuation from the CGM is relevant only if the dominant sources of LyC photons are galaxies similar to the UV color-selected galaxies in our sample ($0.25 < L/L^{*} < 3.0$ at $\langle z \rangle = 2.3$) - for other galaxies, these measurements are an approximation. However, in all cases the general statement holds that \mfp\ and related statistics are affected by the $\sim$Mpc-scale gaseous environments of the sources that contribute significantly to the metagalactic background. 

The IGM and CGM opacities measured in this paper have significant implications for studies of ionizing sources at high redshift and for estimates of the metagalactic UV background at $z\approx2-3$.

\acknowledgements
The authors wish to thank Claude-Andr{\'e}  Faucher-Gigu{\`e}re for his careful reading of the draft and pertinent comments. We also thank Olivera Rakic for her contributions to the reduction of the QSO data set and for her helpful advice. The authors wish to acknowledge Ryan Cooke who contributed the fits to the damped profiles in our QSO spectra. Our thanks to Bob Carswell for his assistance with VPFIT.  

We wish to acknowledge the staff of the the W.M. Keck Observatory whose efforts insure the telescopes and instruments perform reliably. Further, we extend our gratitude to those of Hawaiian ancestry on whose sacred mountain we are privileged to be guests. 

This work has been supported by the US National Science Foundation through grants AST-0606912 and AST- 0908805. CCS acknowledges additional support from the John D. and Catherine T. MacArthur Foundation and the Peter and Patricia Gruber Foundation. This research has made use of the Keck Observatory Archive (KOA), which is operated by the W. M. Keck Observatory and the NASA Exoplanet Science Institute (NExScI), under contract with the National Aeronautics and Space Administration.

\bibliographystyle{apj}
\bibliography{mfp}

\appendix

\section{Fits to high-\NHI\ absorbers}

\label{App_fitting}

As discussed in \S \ref{intro},  for $\log(\NHI / \rm cm^{-2} ) \gtrsim 14.5$, the Ly$\alpha$ transition is saturated, complicating the measurement of absorbers with higher \NHI\ from Ly$\alpha$ only spectra. Similarly, for absorbers with $\log(\NHI / \rm cm^{-2} ) \gtrsim 15.5$, the Ly$\beta$ transition saturates. Thus, the precision and accuracy with which high-\NHI\ systems can be measured will depend to some degree on the number of accessible Lyman lines, thus imposing some redshift dependence. 


For the sample presented in this paper, the QSO spectra cover the Ly$\alpha$ and Ly$\beta$ transition for all the absorbers. Given the typical spectral coverage of our sample (see Table \ref{field}), for absorbers with $z\gtrsim2.3$, the Ly$\gamma$ transition is also covered, and for those with $z\gtrsim2.4$, Ly$\alpha,\beta,\gamma,\delta$ are generally observed. For absorbers with $z\gtrsim2.5$, the spectral coverage of the QSOs typically allows for observation of transitions all the way to the Lyman Limit, thus allowing for accurate measures of \NHI\ to $\log(\NHI/\rm cm^{-2}) \lesssim 17.2$.

 An additional complication arises from the fact that high-\NHI\ absorbers are often found in blended systems of several high-\NHI\ absorbers. Such absorbers, when observed in only Ly$\alpha,\beta$, often cannot be de-blended. The incidence of such blends increases with increasing redshift as the density of the forest increases. This blending effect is mitigated in our sample at $z>2.4$ due to access to addition Lyman series lines, but may go unrecognized in the low end of the redshift interval ($z \lesssim 2.4$) which lack optically thin Lyman lines.
For such blended systems in the presented sample with $z\lesssim2.4$, the number of subcomponents may be underestimated which may lead to some degree of incompleteness in the sample. 

In this appendix we assess the accuracy of our fits to absorbers with $\log(\NHI / \rm cm^{-2} ) \gtrsim 15.5$.

\subsection{Examples of fits to high-\NHI\ absorbers}

\label{App_examples}

\begin{figure}
\center
\includegraphics[width=\textwidth]{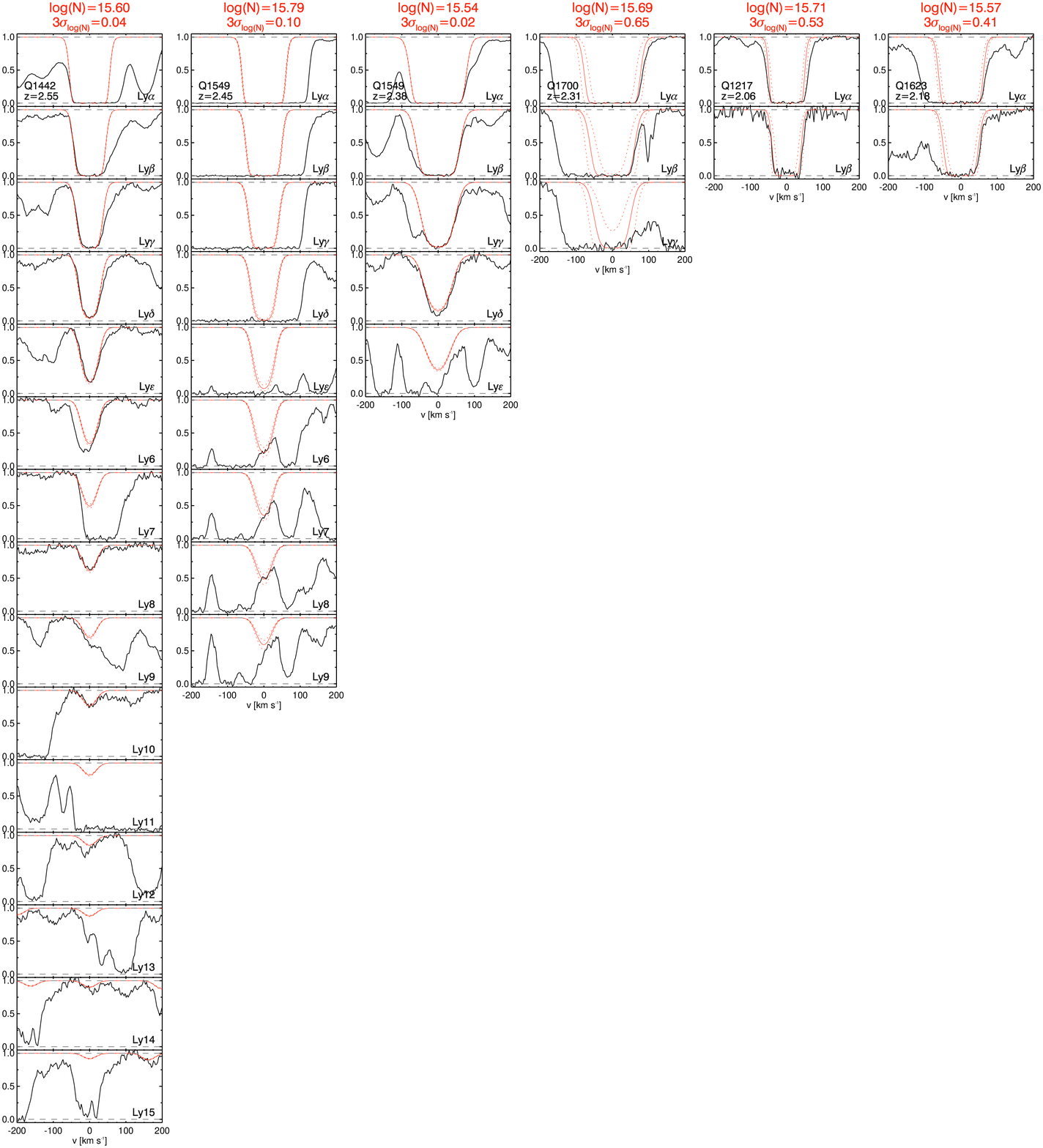}
\caption{\small Example fits to absorbers with $15.5<\log(\NHI/\rm cm^{-2}) <16.0$. The black curves show the HIRES data. Grey dashed lines mark the zero point and continuum of the QSO spectra. The solid red curve shows the best fit Voigt profile to each absorber. The dashed red curves show the formal $\pm3\sigma$ \NHI\ error.}
\label{fit155-160}
\end{figure}

\begin{figure}
\center
\includegraphics[width=0.9\textwidth]{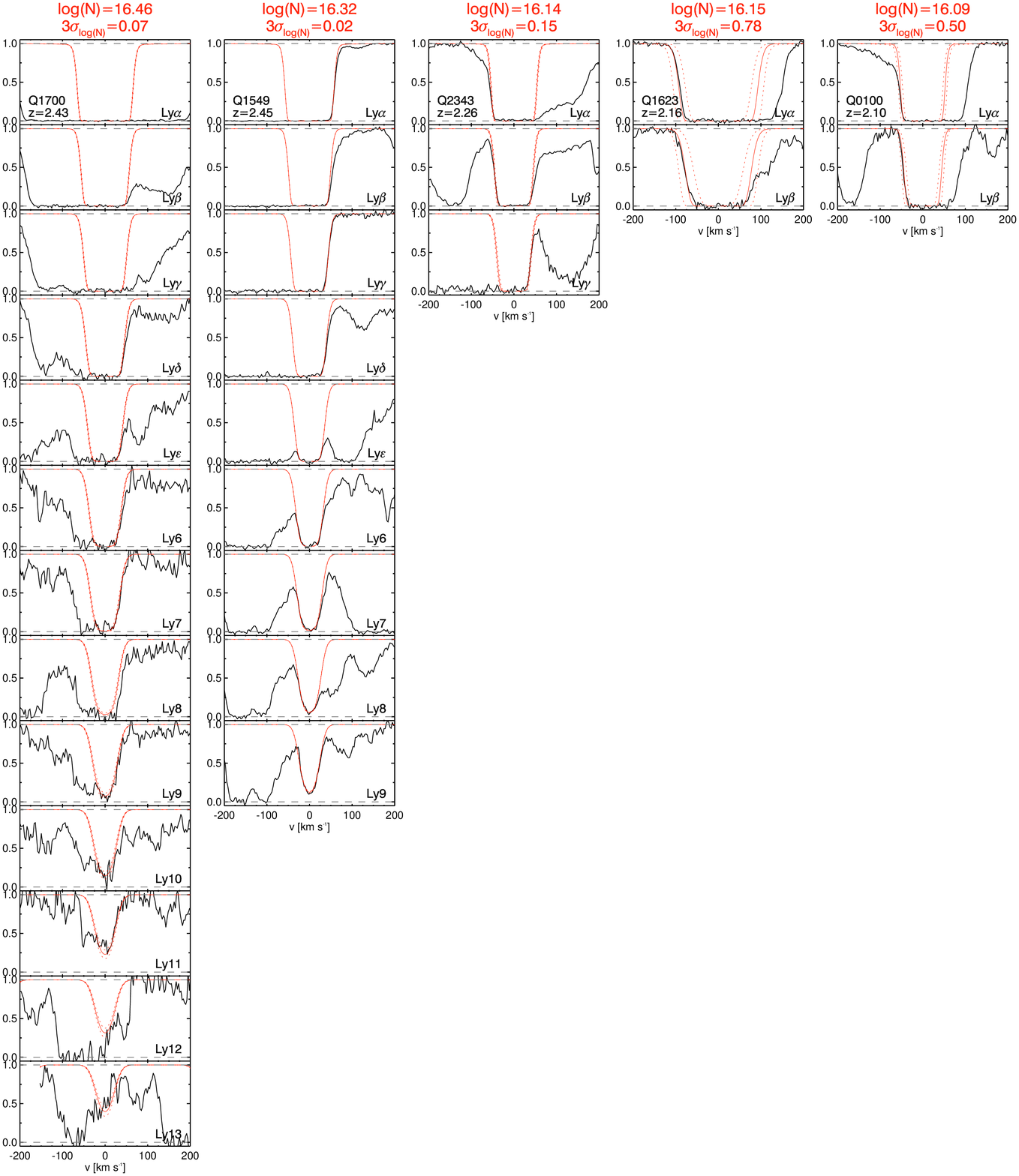}
\caption{\small Example fits to absorbers with $16.0<\log(\NHI/\rm cm^{-2}) <16.5$. The black curves show the HIRES data. Grey dashed lines mark the zero point and continuum of the QSO spectra. The solid red curve shows the best fit Voigt profile to each absorber. The dashed red curves show the formal $\pm3\sigma$ \NHI\ error.}
\label{fit160-165}
\end{figure}

\begin{figure}
\center
\includegraphics[width=0.9\textwidth]{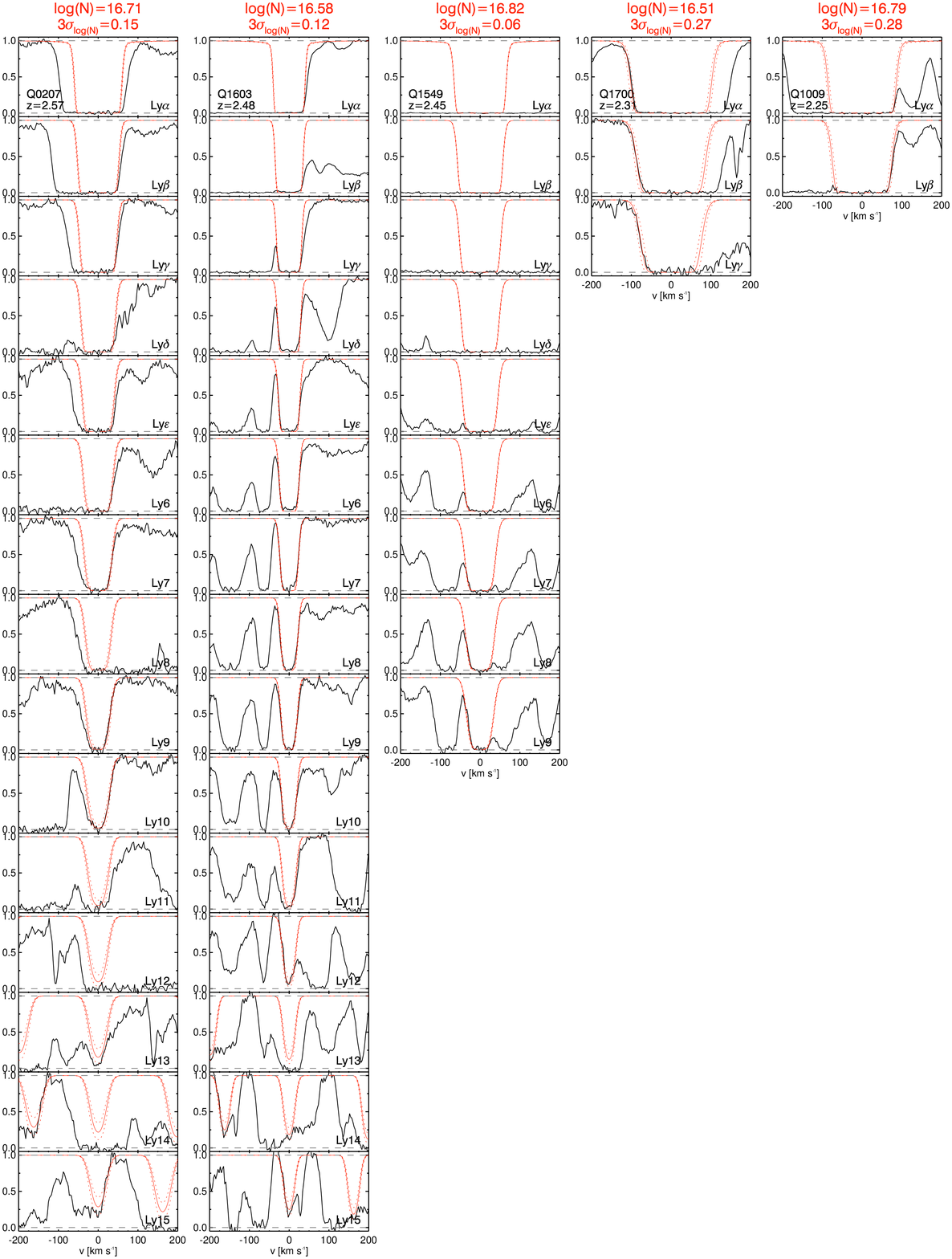}
\caption{\small Example fits to absorbers with $16.5<\log(\NHI/\rm cm^{-2}) <17.0$. The black curves show the HIRES data. Grey dashed lines mark the zero point and continuum of the QSO spectra. The solid red curve shows the best fit Voigt profile to each absorber. The dashed red curves show the formal $\pm3\sigma$ \NHI\ error.}
\label{fit165-170}
\end{figure}

Figures \ref{fit155-160} - \ref{fit165-170} show examples of the fits to the HIRES spectra for absorbers with $15.5<(\log(\NHI/ \rm cm^{-2})<17.0$ . In each case, the HIRES data are shown in black, and the solid red curve shows the model with the best fit to the data. The dashed curves show the $\pm3\sigma$ formal errors on \NHI\ as reported by VPFIT. As expected, the errors on \NHI\ are dependent on the number of transitions observed; however, the dependence is not trivial as the effect of (a) blending from other proximate absorbers, (b) the contamination of higher-order Lyman transitions due to unrelated absorbers, and (c) the S/N of the spectrum in the areas of interest all affect our ability to accurately determine the parameters of the absorber. 

\subsection{Testing fits with only Ly$\alpha$ and Ly$\beta$}

\label{App_test}

To assess the accuracy of fits to absorbers with $\log(\NHI / \rm cm^{-2} ) \gtrsim 15.5$ observed only in Ly$\alpha$ and Ly$\beta$, we use the set of absorbers with $\log(\NHI / \rm cm^{-2} ) \gtrsim 15.5$ for which $n\ge5$ Lyman transitions were measured, generally drawn from the subsample with $z>2.4$. We then attempt to refit these absorbers using VPFIT with constraints from the Ly$\alpha$ and Ly$\beta$ region only. The outcomes of these tests can be generalized into two principal categories: highly-blended and less-blended absorbers.


Absorbers for which a portion of one of both edges/wings of the absorption line can be observed in Ly$\alpha$ and/or Ly$\beta$ can be fit much more reliably. Such systems, when refit with VPFIT using only their Ly$\alpha$ and Ly$\beta$ transition result in weaker constraints on the line parameters as expected, but result in \NHI\ determinations consistent within their formal errors with those obtained via the higher-order transitions. The fits are not systematically offset to higher or lower values of \NHI, and therefor should not \textit{bias} the measurements of $f(N,X)$ in any large way. 

For absorbers whose Ly$\alpha$ and Ly$\beta$ transition are fully saturated and both edges/wings of the absorption line cannot be observed due to other neighboring absorbers (e.g. see the left-most panel in Figure \ref{fit160-165} and the center panel in Figure \ref{fit165-170}), essentially no information is available on the sub-component structure of the system. 
Such systems are often fit with a minimum of two components because it is rare that the wings/edges of the blended system are symmetric. In the absence of higher-order Lyman lines, the number of subcomponents is typically underestimated, likely leading to some degree of incompleteness in the lower-redshift portion of the catalog. There would also be a tendency to underestimate the total \NHI\ in such blends. If the \NHI\ of individual absorbers in the blends were systematically underestimated, this would \textit{bias} the measurement of $f(N,X)$ to have steeper values of $\beta$ than the true value. 

In conclusion, the lack of higher-order constraints on the lower-$z$ saturated absorbers have two principal effects on our measurements. (1) Relatively unblended absorbers yield somewhat weaker constraints on the parameters that are unlikely to bias the measurement of $f(N,X)$. (2) Highly-blended absorbers yield very weak constraints from Ly$\alpha$ and Ly$\beta$ only and therefore are typically fit with fewer subcomponents and with lower-\NHI\ values than would be assigned using additional Lyman series constraints. This results in some degree of incompleteness in the lower-$z$ portion of the catalog. 

\subsection{Assessing Incompleteness in the $z\lesssim2.4$, $log(\NHI/\rm cm^{-2}) > 15.5$ Catalog}

\label{incomplete}

To estimate the degree of potential incompleteness in the lower-$z$ portion of the catalog due to line blending, we consider the fraction of the higher-$z$, high-\NHI\ absorbers observed in $n\ge5$ Lyman series transitions
 that would be unconstrained or miscounted with observations of Ly$\alpha$ and Ly$\beta$ alone. This method likely overestimates the incompleteness as the degree of line blending and contamination is higher in the high-redshift forest than in the lower-$z$ forest due to evolution in the line density as a function of redshift. Nevertheless, we report the results of this test, considering separately absorbers in half-dex bins of \NHI.  The results are summarized in Table \ref{incomp_table}.

 \begin{deluxetable}{lcccccccc}
\tablecaption{Constraints on \NHI\ vs. the number of observed Lyman series transitions}
\tablewidth{0pt}
\tablehead{
\colhead{$\log(\NHI/\rm cm^{-2}$)} &  \colhead{Lines considered } & \colhead{\# with 5+ } &  \multicolumn{2}{c}{strongly constrained} & \multicolumn{2}{c}{weakly constrained\tablenotemark{b}} & \multicolumn{2}{c}{unconstrained\tablenotemark{c} } \\
\colhead{} & \colhead{in the fit} & \colhead{Lyman series lines\tablenotemark{a} } & \colhead{\#} & \colhead{\%} & \colhead{\#} & \colhead{\%} & \colhead{\#} & \colhead{\%} 
}
\startdata
15.5 - 16.0 &  Ly$\alpha,\beta$~~~~ & 23 & 14 & 61\% & 7
& 30\% & 2 &~9\% \\
15.5 - 16.0 & Ly$\alpha,\beta,\gamma$ & 23 & 18 & 78\% & 3 & 13\% & 2 & ~9\% \\
16.0 - 16.5 &  Ly$\alpha,\beta$~~~~ & 12 & 2 & 17\% & 2 & 17\% & 8 & 67\%  \\
16.0 - 16.5 & Ly$\alpha,\beta,\gamma$ &  12 & 2 & 17\% & 6 & 50\% & 4 &33\% \\
16.5 - 17.0 &  Ly$\alpha,\beta$~~~~ & ~5 & 1 & 20\% & 2 & 40\% & 2 & 40\% \\
16.5 - 17.0 &  Ly$\alpha,\beta,\gamma$ & ~5 & 3 & 60\% & 0 & ~0\% & 2 & 40\%
 \enddata
	\tablenotetext{a}{The number of absorbers in the catalog with \NHI\ in the range listed and at least 5 Lyman series transitions observed.}
	\tablenotetext{b}{For weakly constrained absorbers, the transitions listed give unambiguous evidence of the presence of an absorber, and typically have at minimum one edge of one transition well observed without a significant blend. }
	\tablenotetext{c}{For unconstrained absorbers, the transitions listed are fully saturated and blended with other high-\NHI\ lines (e.g. see middle panel Figure \ref{fit165-170}) meaning this absorber would likely not be counted or be very poorly fit in data with only the transitions listed.  }
     \label{incomp_table}
\end{deluxetable}

For absorbers with $15.5<\log(\NHI/\rm cm^{-2}) <16.0$, 23 are measured with 5 or more Lyman series transitions. Of these, only 2 would have essentially no constraint on the parameters of the absorber from Ly$\alpha$ and Ly$\beta$ alone (e.g. Figure \ref{fit155-160}, second from the left). For 7/23 of these systems, the constraint on \NHI\ would be relatively weak, however, an individual component associated with that absorber can be clearly identified. All 7 of these systems are fit with much improved accuracy with the additional constraint from Ly$\gamma$.
Among absorbers with $16.0<\log(\NHI/\rm cm^{-2}) <16.5$, 
we observed 12 absorbers with constraints from 5 or more Lyman series transitions. Of these, only 4 are strongly constrained from Ly$\alpha$ and Ly$\beta$ alone. With Ly$\gamma$, 8/12 could be fit, and with Ly$\alpha,\beta,\gamma,\delta$, 11/12 would be strongly constrained and could accurately be fit. 
For absorbers with \NHI\ just below that of LLSs ($16.5<\log(\NHI/\rm cm^{-2}) <17.0$), 
we observe 5 systems in 5 or more Lyman series transitions.\footnote{All five absorbers are observed in a minimum of 9 transitions. We do not observe any absorbers in this \NHI\ range with 5-8 Lyman series transitions observed.} Of these, only 2 would have no constraint on \NHI\ from Ly$\alpha$ and Ly$\beta$ alone. 

From this assessment based on higher-$z$ measurements, we expect that the principal bias in our measurements of \NHI\ for high-\NHI\ systems for absorbers with only Ly$\alpha$ and Ly$\beta$ is to under-count the number of systems with $\log(\NHI/\rm cm^{-2}) > 15.5$. Within the present sample, systems with $16.0<\log(\NHI/\rm cm^{-2}) <16.5$ appear to be most-affected by blending and may therefore be the most incomplete. In the next appendix, we add further evidence to this claim by comparing the measurement of $f(N,X)$ for absorbers in two redshift subsamples. 

\section{The Effect of the accuracy of $f(N,X)$ on the measurement of the \mfp}

\begin{figure}
\center
\includegraphics[width=0.5\textwidth]{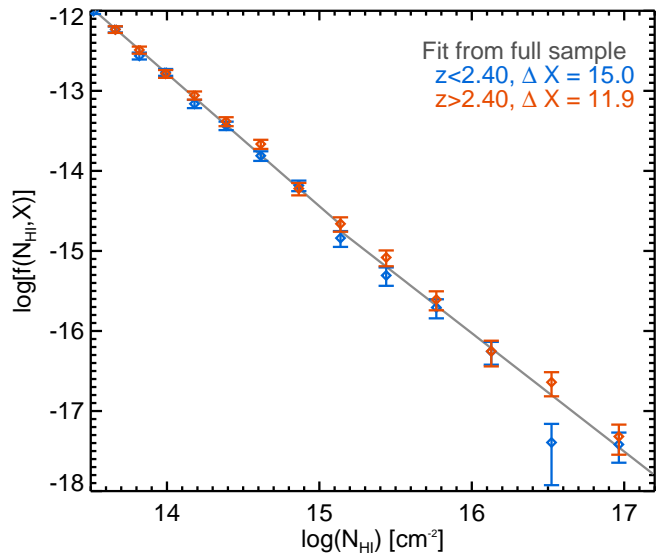}
\caption{\small The frequency distribution measured from the higher-redshift (red) and lower-redshift (blue) portions of the absorber sample (many of the points from the two subsamples lie on top of each other). The grey curve shows the broken power law fit to the full absorbers sample from \S \ref{broken_text}.  }
\label{fNX_z}
\end{figure}

\label{App_fN}


Here we assess the redshift dependence of the frequency distribution presented in \S \ref{fN_text} and \S \ref{LyC}. We split the absorber sample roughly in half and compare those absorbers with $z<2.4$ and those with $z>2.4$. For the higher-redshift subsample, the spectra typically cover at least 4 Lyman series transitions and most of this sample have all the Lyman series transitions observed.  Figure \ref{fNX_z} shows the results of this comparison. Notably, the higher-redshift subsample (red points) is in reasonable agreement with the broken power law fit (gray curves) presented in \S \ref{broken_text} that were used in the Monte Carlo simulation in \S \ref{mcText}. Further, the higher-redshift sample, tends to lie slightly above the fit. Some portion of this is likely due to the redshift evolution of the line density of the forest; however, it is also possible that $f(N,X)$ of absorbers with $\log(\NHI/\rm cm^{-2}) >15.5$ is actually higher than that measured from the full sample. As discussed in Appendix \ref{incomplete}, we expect that the lower-$z$ catalog likely suffers from incompleteness due to line blending, especially for absorbers with $16.0<\log(\NHI/\rm cm^{-2}) <16.5$. In Figure \ref{fNX_z}, we note the only bin for which the low-$z$ and high-$z$ data do not agree within $1\sigma$ contains absorbers with $16.0<\log(\NHI/\rm cm^{-2}) <16.5$.

The principal result of this paper is the relatively high measured value of $f(N,X)$ for absorbers with $14.5<(\log(\NHI/ \rm cm^{-2})<17.2$ which results in a value of  \mfp\ comparable to or smaller than previous measurements. If we were to use the high-redshift sub-sample from Figure \ref{fNX_z} to measure the opacity and \mfp, we would find higher IGM opacity and therefore a shorter (more discrepant) \mfp. In any case, if there is a systematic bias in the measurements, it is one that would conspire to make \mfp\ appear \textit{larger} than its true value. We therefore conclude that it is unlikely that a bias in the precision of our measurements of $f(N,X)$ as a function of $z$ would change the principal result of the paper.

\end{document}